\documentclass[preprint,review,12pt]{elsarticle}

\usepackage{lineno,hyperref}
\modulolinenumbers[5]

\journal{Combustion and Flame}

\usepackage{amsmath}
\usepackage{epstopdf}
\usepackage{graphicx}
\usepackage{color}
\usepackage{soul}

 \pdfoutput=1 
\usepackage{geometry}
\usepackage{setspace} 

\usepackage{amssymb}
\usepackage{amsmath}
\usepackage{multicol}
\usepackage{latexsym}
\usepackage{subfigure}
\usepackage{multirow}
\usepackage{setspace}
\graphicspath{{Pictures/}}

\usepackage{longtable}
\usepackage[labelsep=period]{caption}
\biboptions{sort&compress}

\usepackage[table]{xcolor}

\usepackage{multirow}
\usepackage{arydshln}
\usepackage{amsmath,amssymb}

\usepackage{tikz}
\usetikzlibrary{calc,topaths}
\usepackage{notoccite}

\usepackage{setspace} 
\usepackage{epstopdf}

\makeatletter
\def\ignorecitefornumbering#1{%
	\begingroup
	\@fileswfalse
	#1
	\endgroup
}
\makeatother

\begin{document}

\begin{frontmatter}



\title{\LARGE {Physics-informed data-driven prediction of Jet A-1 spray characteristics using time-resolved flame chemiluminescence and sparse Mie scattering}}

\author{{\large Liam Krebbers, Sajjad Mohammadnejad, Ali Rostami, and Sina Kheirkhah$^{*}$\\[10pt]
		{\footnotesize \em School of Engineering, The University of British Columbia, Kelowna, Canada, V1V 1V7}\\[0pt]}}
	
\date{}


\cortext[cor1]{Corresponding author: sina.kheirkhah@ubc.ca}

\begin{abstract}
A time-lag and linear regression-based framework is developed and its performance is assessed for predicting temporally resolved spray number of droplets using flame chemiluminescence and a sparse number of droplets data. Separate pressure, interferometric laser imaging for droplet sizing, shadowgraphy, and flame chemiluminescence are performed for the spray characterization. Simultaneous 10~kHz flame chemiluminescence and 0.2~Hz Mie scattering measurements are performed for the purposes of the framework development and the number of droplets prediction. Both methane and/or Jet A-1 are used in the experiments. Three conditions corresponding to perfectly premixed methane and air, Jet A-1 spray, and Jet A-1 spray in premixed methane and air flames are examined. For all test conditions, the fuels and air flow rates are adjusted to produce a fixed power of 10~kW. The results show that the frequency of the spatially averaged flame chemiluminescence as well as the number and the mass of the droplets (for both reacting and non-reacting conditions) oscillations frequencies match; however, these frequencies do not match that of the pressure fluctuations. This suggests that the flame chemiluminescence dynamics is driven by the fuel injection system. For signals with matching frequency content, a data-driven framework is developed for predicting an objective signal (the spray number of droplets) using an input signal (the flame chemiluminescence). The performance of the developed framework is assessed for tested spray conditions and the predicted number of droplets agrees well with those measured. For gas turbine engine combustion research, the developed framework is of importance, as it facilitates understanding the time-resolved spray characteristics for instances that the spray data is available sparsely.

\end{abstract}


\end{frontmatter}

\section{Introduction}
\label{Int}
The \textit{modus operandi} of existing civil aviation engines is turbulent spray combustion. Spray flames feature complex interactions between turbulence, liquid fuel atomization and transport, as well as combustion chemistry. Such complex interactions allow for the presence of several pathways for thermoacoustic coupling~\cite{lieuwen2005combustion,o2015transverse,mcmanus1993review,ducruix2003combustion}, which are detrimental to the engine operation and can lead to poor combustion emissions and, sometimes, system failure~\cite{lieuwen2003modeling}. Despite several thermoacoustic-related investigations have been performed in the past decades and many review papers have been published (see for example~\cite{mcmanus1993review,lieuwen2003modeling,ducruix2003combustion,dowling2005feedback,vignat2020combustion,o2022understanding}), our understanding of the coupling between sprays and their flames remains to be further developed. This is challenging, as numerous thermo-fluidic parameters are required to be measured with large spatio-temporal resolutions. This challenge is further accentuated for multi-nozzle configurations~\cite{prieur2017ignition,kim2022experimental,topperwien2022analysis} and/or at high-pressure conditions~\cite{philo2021100,kheirkhah2017dynamics,skeen2015simultaneous}. Acknowledging the importance of the technical efforts made to address the above experimental challenges, data-driven approaches may be developed and implemented to predict information that is difficult to acquire or missing from the experiments. The present study is motivated by the need for developing and assessing a data-driven framework that allows for predicting the spray flames characteristics.

For perfectly premixed flames, the pressure oscillations inside the combustion chamber can lead to the oscillations of the injected fuel and air mixture, which is usually followed by vortex shedding or deformation of a helical vortex structure inside the combustion chamber~\cite{poinsot1987vortex,schadow1989large,caux2014thermo}. This is accompanied by periodic variations of the flame surface density, and as a result, periodic heat release rate oscillations~\cite{steinberg2010flow}. For technically-premixed flames (or partially premixed flames), the thermoacoustic coupling may occur due to the periodic spatial and temporal variations of the fuel-air equivalence ratio~\cite{lieuwen1998role,lee2000measurement,sattelmayer2003influence,kather2021interaction}. In addition to the above thermo-fluidic pathways, the structural oscillations can also create a feedback mechanism for thermoacoustic coupling of perfectly and technically premixed flames, as discussed in~\cite{zhang2019numerical,shahi2018strongly,heydarlaki2022competing}.

The thermoacoustic coupling can be quantified using the Rayleigh gain, which requires time/phase resolved information related to the pressure and the heat release rate oscillations~\cite{durox2009rayleigh,selimefendigil2011nonlinear,chen2020influence,magri2020sensitivity}. Of importance is measurement of the heat release rate; and, for perfectly premixed and atmospheric flames, experimental observables such as the planar laser-induced fluorescence of $\mathrm{OH} \times \mathrm{CH_2O}$, $\mathrm{H} \times \mathrm{CH_2O}$, $\mathrm{HCO}$, and $\mathrm{CH}$ may be used for quantifying the heat release rate~\cite{najm1998adequacy,wabel2017measurements,skiba2018premixed,zhou2017thin,mohammadnejad2019internal,mohammadnejad2021contributions,mohammadnejad2022hydrogen}. For atmospheric Jet A-1 spray flames subjected to self-excited thermoacoustic oscillations, Apeloig \textit{et al.}~\cite{apeloig2015liquid} performed high-speed simultaneous pressure and planar laser-induced fluorescence of both Kerosene and OH. They~\cite{apeloig2015liquid} utilized an air-blast injector that created a liquid film, which was atomized and carried into the combustion chamber. Their results showed that the air flow rate fluctuations alter the trajectory of the atomized droplets, and this alteration created a pulsation in the spray. For a high-pressure combustor, Kheirkhah \textit{et al.}~\cite{kheirkhah2017dynamics} employed high-speed and times-resolved pressure, flame chemiluminescence, and Stereoscopic-Particle Image Velocimetry. Their results~\cite{kheirkhah2017dynamics} showed that the spray velocity features fluctuations at dominant frequencies similar to those of the pressure and heat release rate. Recently, Passarelli \textit{et al.}~\cite{passarelli2021cross} studied self-excited oscillations of Jet A-1 at high pressure. Their results showed that, while the flame chemiluminescence and Jet A-1 spray may feature large amplitude fluctuations at a matching frequency, the dominant frequency of pressure oscillations may be different than those of the spray and flame chemiluminescence.



Although past experimental studies related to the themroacoustics of spray flames are of significant importance as they elaborate the underlying coupling mechanism in combustors operating with liquid fuels, the spray is often studied qualitatively. For example, the spatially integrated fuel droplets Mie scattering signal is used \cite{kheirkhah2017dynamics} to understand the spray dynamics. Although such qualitative characterization of the spray is important for understanding the thermoacoustic coupling in liquid fueled combustors, alternative experimental observables, such as the number of droplets and/or the liquid fuel mass inside an illuminated volume/plane, may be used to study the spray characteristics quantitatively. The objective of the present study is to quantify the temporal variation of the spray number of droplets inside a plane using a data-driven framework. Specifically, we aim to utilize time-resolved flame chemiluminescence and sparse Mie scattering data along with a data-driven framework (which is developed here) to predict the variation of the spray number of droplets with time. In the following, the experimental methodology, the spray flames characteristics, the prediction framework, and the results are presented in sections~\ref{EM}--\ref{Results}, respectively. The conclusions are summarized in section~\ref{Conclusions}.

\section{Experimental methodology}
\label{EM}
Details of the utilized experimental setup, diagnostics, and data reduction are elaborated in this section.

\subsection{Experimental setup}
\begin{figure*}[!t]
	\centering
	\includegraphics[width = 1.0\textwidth]{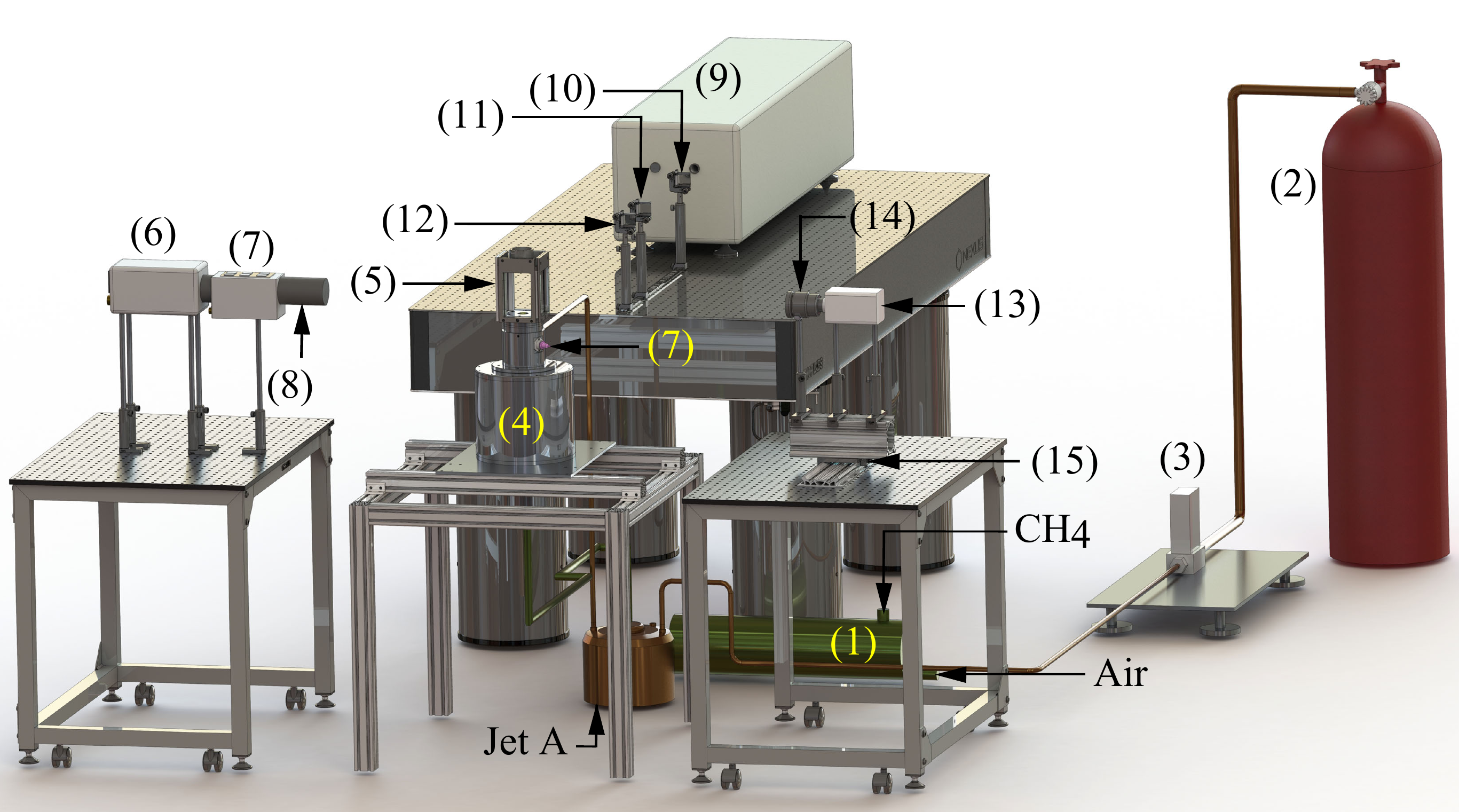}
	\caption{The layout of the utilized experimental setup and diagnostics. Items 1--3 are a mixing chamber, a nitrogen bottle, and a pressure controller, respectively. Items 4 and 5 are the flow conditioning equipment and the gas turbine model combustor, respectively. Items 6--8 are a camera, an intensifier, and a UV lens equipped with a bandpass filter, respectively. Items 9, 10-12, 13, and 14 are a laser, sheet forming optics, a camera, and a lens, respectively. Item 15 is a high-precision rotational stage.}
	\label{fig:diagnostics}
\end{figure*}

The experimental setup refers to the fuel and air delivery system as well as the utilized burner. Air was provided by an Atlas Copco compressor; and, the air flow rate was controlled using an Alicat 5000 MCRH. A gaseous fuel (methane) and a liquid fuel (Jet A-1) were utilized in the present study. Grade 2.0 methane (99$\%$ chemical purity) was provided from a pressurized bottle; and, its flow rate was controlled using a Brooks SLA5853. As demonstrated in Fig.~\ref{fig:diagnostics}, both air and methane were fed into a mixing chamber (see the green cylinder, item 1). The utilized Jet A-1 density at standard pressure and temperature conditions was measured and equals 812.0~$\mathrm{kg/m^3}$. The minimum flash point, the freezing point, and the distillation end point of the utilized Jet A-1 are 38, -47, and 300$^\mathrm{o}\mathrm{C}$, respectively. The maximum aromatic and sulfur concentrations of the utilized Jet A-1 are 25\% by volume and 0.3\% by mass, respectively. The above temperatures and concentrations were provided by the fuel producer. Grade 5.0 (99.999\% chemical purity) nitrogen was purged into a pressurized vessel that carried Jet A-1, using the bottle shown as item 2 in Fig.~\ref{fig:diagnostics}. During each experiment, the pressure of nitrogen in the fuel vessel was fixed using an Alicat dual-valve pressure controller (see item 3 in Fig.~\ref{fig:diagnostics}). The spray flow rate was calibrated, with details of the calibration procedure provided in Appendix~A.

\begin{figure*}[!t]
	\centering
	\includegraphics[width = 0.9\textwidth]{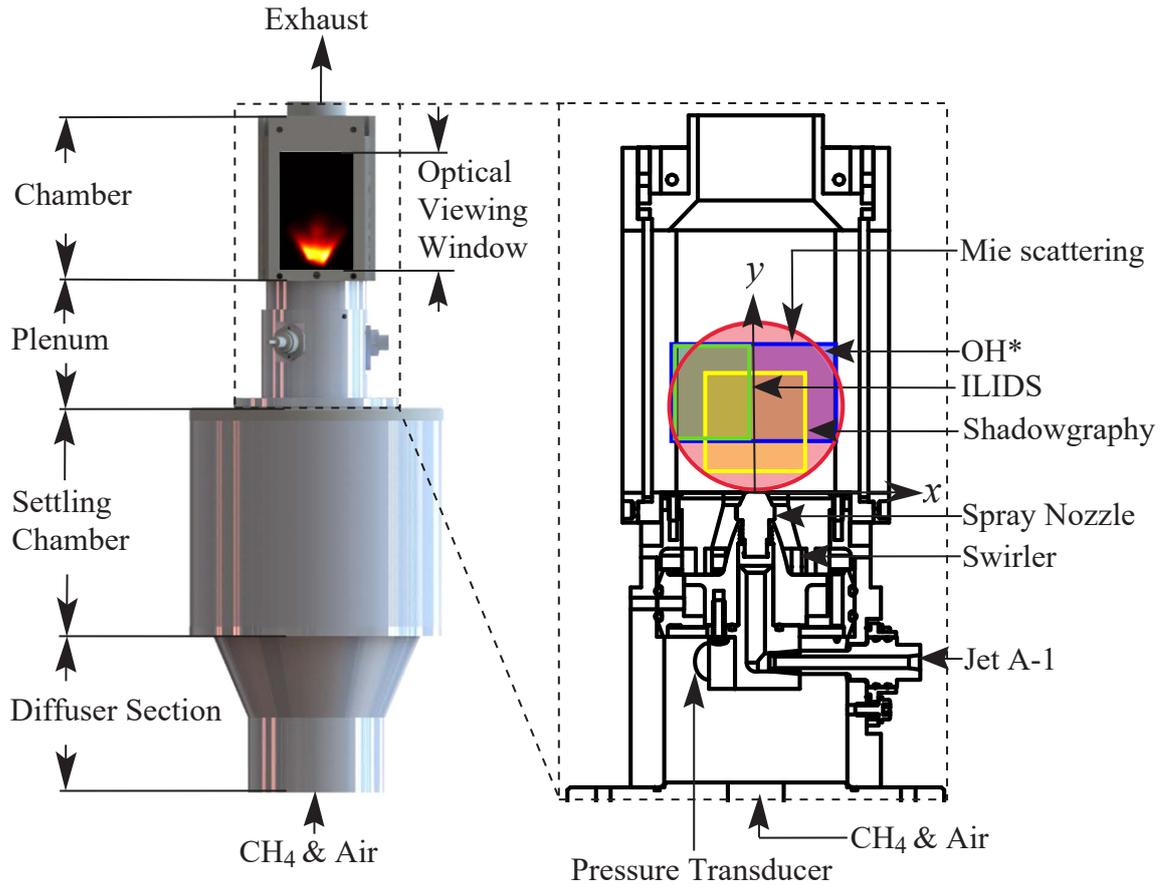}
	\caption{The burner, which is composed of a diffuser section, a settling chamber, a plenum, and a combustion chamber. The inset on the right-hand-side presents the combustor, which includes the plenum and the combustion chamber. The Mie scattering, Shadowgraphy, and ILIDS fields of view are shown by the blue, yellow, and green squares, respectively. The flame chemiluminescence field of view is shown by the red circle.}
	\label{fig:burner}
\end{figure*}

The mixture of methane and air was provided at the bottom to the flow conditioning equipment, see item~4 in Fig.~\ref{fig:diagnostics}. For clarity, the flow conditioning equipment and the combustor details are presented in Fig.~\ref{fig:burner}. As shown in the figure, the mixture of methane and air enters a diffuser section (with an area ratio of 4:1). This is followed by a settling chamber, which is equipped with 5 equally spaced mesh screens. The details of the diffuser section and the settling chamber are identical to those used in~\cite{mohammadnejad2022new,saca2022preheat,mohammadnejad2020thick}. Downstream of the settling chamber, a gas turbine model combustor, which includes a plenum and a combustion chamber are installed, as shown in Fig.~\ref{fig:burner}. The plenum and the chamber are originally designed by Turbomeca and are similar to those presented in Weigand \textit{et al.}~\cite{weigand2006laser} for gaseous fuels combustion. In Weigand \textit{et al.}~\cite{weigand2006laser}, the plenum carries a conical bluff-body; however, in the present study and to accommodate for the spray injection, similar to Wang \textit{et al.}~\cite{wang2019soot}, the conical bluff-body was replaced by a Delavan pressure swirl atomizer (see Fig.~\ref{fig:burner}). The combustion chamber is equipped with 4 fused silica windows for optical accessibility. A Cartesian coordinate system was used in the present study. The origin of the coordinate system is at the exit plane of the spray injector and the injector centerline. The $y$--axis of the coordinate system coincides with the chamber centerline; and, the $x$--axis is normal to the $y$--axis and the combustion chamber side walls.

\subsection{Diagnostics}
\begin{figure*}[!t]
	\centering
	\includegraphics[width = 1.0\textwidth]{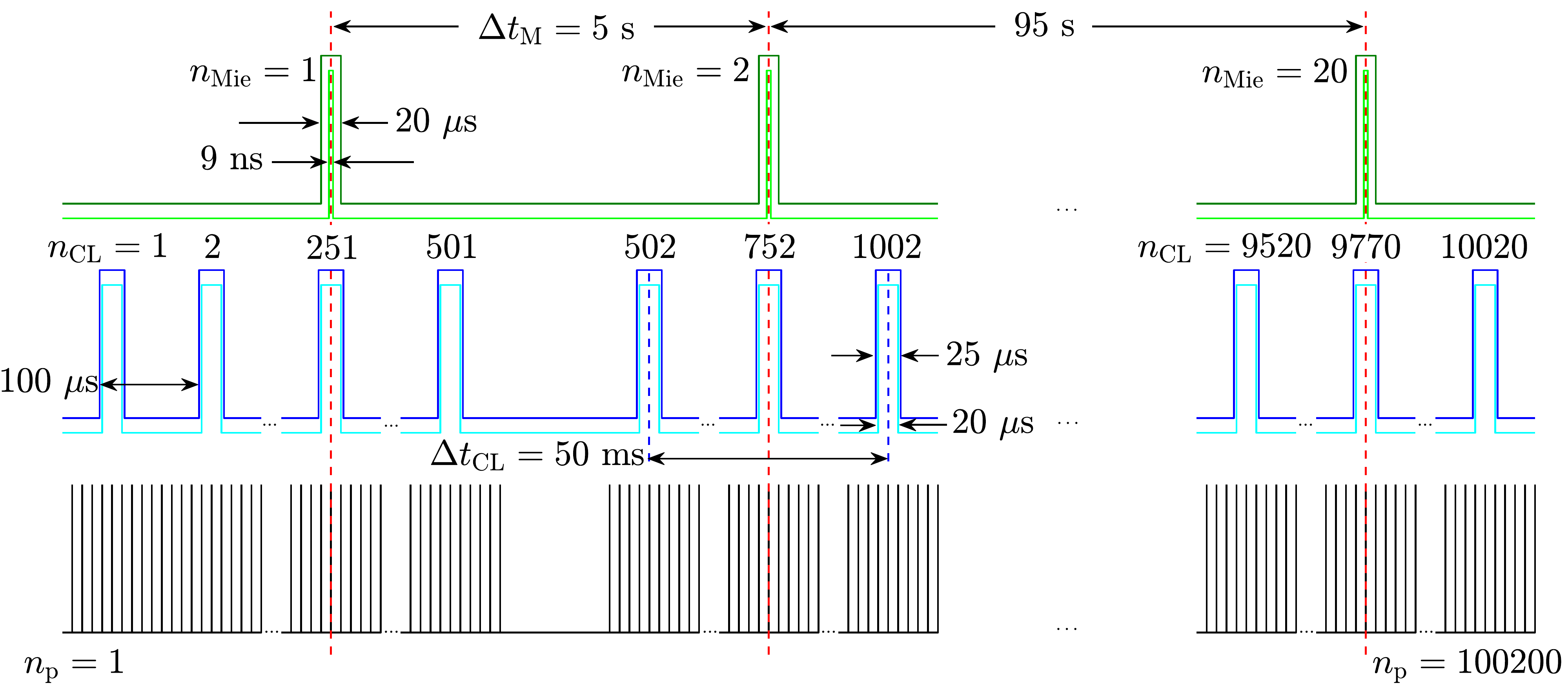}
	\caption{The timing between the synchronized Mie scattering, flame chemiluminescence, and pressure measurements. The light and dark green lines present the laser excitation and camera exposure timing, respectively, for the Mie scattering measurements. The light and dark blue lines present the timing signal for the intensifier gate and the chemiluminescence camera exposure, respectively. The black lines represent the timing for the pressure measurements.}
	\label{fig:timing}
\end{figure*}

Simultaneous plenum pressure, high-speed flame chemiluminescence, and low-speed Mie scattering data were collected for the purposes of developing and testing the framework of the present study. For the above simultaneous measurements, the acquisition frequency of pressure, flame chemiluminescence, and Mie scattering measurements are 100000, 10000, and 0.2~Hz respectively. For each test condition (elaborated later in this section), 20 datasets of pressure, flame chemiluminescence, and Mie scattering were collected every 5~s, with the details of the acquisition timing shown in Fig.~\ref{fig:timing}. In the figure, $n_\mathrm{Mie}$, $n_\mathrm{CL}$, and $n_\mathrm{p}$ correspond to the number of the collected Mie scattering image, flame chemiluminescence image, and pressure data, respectively. As shown by either of the red dashed lines in the figure, the Mie scattering image (see the light and dark green signals) is simultaneously acquired with one chemiluminescence image (see the light and dark blue signals) and one pressure data (see the black signal) for each dataset. Additionally, 250 chemiluminescence images before as well as 250 chemiluminescence images after the above simultaneously collected chemiluminescence image is acquired, see Fig.~\ref{fig:timing}. Also, 4 pressure data before and 5 pressure data after the above simultaneously collected pressure data was acquired. It is important to highlight that, for each dataset, the flame chemiluminescence and pressure measurements are time-resolved; however, the Mie scattering measurement is not. 

In addition to the above simultaneous measurements, separate pressure, flame chemiluminescence, shadowgraphy, and Interferometric Laser Imaging for Droplet Sizing (ILIDS) were also performed to characterize the spray flames. The data acquisition frequency for the separate pressure, chemiluminescence, shadowgraphy, and ILIDS were 100000, 10000, 10000, and 5 Hz, with the data collection durations of 100, 0.5, 0.5, and 100~s, respectively. For the purposes of spectral analysis, the separate flame chemiluminescence and shadowgraphy measurements were repeated 6 times. Further details regarding the above diagnostics and the reduction of the collected data are presented in the following.

\subsubsection{Pressure measurements}
A pressure transducer (Model 106B52 from PCB Piezotronics) was installed flush mount with the plenum wall, see Fig.~\ref{fig:burner}. A National Instruments PCIe 6361 data acquisition system was used to collect the voltage generated from the pressure transducer. The transducer features a sensitivity of 716.3 mV/kPa, which was used to convert the collected voltage to pressure data.

\subsubsection{Flame chemiluminescence}
The hardware for acquiring the flame chemiluminescence includes a high-speed camera (Photron Fastcam Nova S12, item 6 in Fig.~\ref{fig:diagnostics}) and a high-speed image intensifier (Invisible vision UVi 1850B, item~7 in Fig.~\ref{fig:diagnostics}). A UV Nikon lens (focal length of 105~mm and aperture number of 1.4) was equipped with a bandpass filter (center wavelength of 310~nm and bandpass width of 20~nm) and was mounted on the intensifier. The lens and the bandpass filter are shown as item~8 in Fig.~\ref{fig:diagnostics}. Although the center wavelength of the bandpass filter is close to the $\mathrm{OH^*}$ emission wavelength, the collected signal from items 6--8 has contributions from emissions of $\mathrm{CO_2^*}$ as well as $\mathrm{C_2^*}$. Passarelli \textit{et al.}~\cite{passarelli2021cross} indicated while the chemiluminescence signal collected near the $\mathrm{CH^*}$ band featured significant contributions from broadband emissions (such as $\mathrm{C_2^*}$), the chemiluminescence signal collected near the $\mathrm{OH^*}$ emission band (similar to the present study) did not feature significant contributions from the broadband emissions in their investigation. Nonetheless, in the present study, the collected chemiluminescence signal is not deemed as an accurate indicator of the heat release rate, may only qualitatively relate to the exothermic processes inside the combustion chamber, and may be used for understanding the flame dynamics, similar to those in~\cite{kheirkhah2017non,kheirkhah2017dynamics}. For both separate and simultaneous flame chemiluminescence measurements reported here, the camera exposure time and the intensifier gate were set to 25 and 20~$\mu$s, respectively, as shown in Fig.~\ref{fig:timing}. The intensifier gain was set to 40\% for all measurements.

The raw chemiluminescence images were subtracted by the mean of 10000 background images. The background subtracted chemiluminescence images were normalized by a Whitefield, which was collected using an NL-360ARC Neewer LED lamp. Then, the images were denoised using an $11 \times 11$ median-based filter. The pixel size for the chemiluminescence images was 68.4~$\mu$m. However, the effective spatial resolution was determined using the USAF~1951 target plate; and, this resolution was 198.4~$\mu$m as discussed in Appendix~B. The flame chemiluminescence field of view was a circle, with a diameter of 75.5~mm. Its center was positioned at $x = 0$ and $y=38.0$~mm, see Fig.~\ref{fig:burner}. Further details regarding the flame chemiluminescence hardware and the data reduction procedure can be found in~\cite{mosadegh2022role,heydarlaki2022competing}.

\subsubsection{Mie scattering}
\label{subsubsection:mie}
The Mie scattering hardware includes an Nd:YAG laser (Lab-Series-170 laser from Spectra Physics, item~9 in Fig.~\ref{fig:diagnostics}), sheet forming optics (10--12), as well as a camera and its collection optics (items~13 and 14). The beam produced by the laser has a wavelength of 1064~nm, which was converted by a second harmonic generator to a 532~nm wavelength and an 8~mm in diameter beam. The beam energy was measured using model QE25LP-S-MB-QED-INT-D0 from Gentec Electro-Optics. At maximum power, the mean and standard deviation of the 532~nm laser beam energy were about 457.2 and 5.6 mJ per pulse. In order to avoid saturation and maximize the quality of the collected Mie scattering image, the laser was operated at 1\% of its maximum power. The sheet forming optics included a plano-concave cylindrical lens (item~10 in Fig.~\ref{fig:diagnostics} with a focal length of -100~mm, a plano-convex cylindrical lens with a focal length of 500~mm (item~11), and a plano-convex cylindrical lens with a focal length of 1000~mm (item~12). The above optics facilitated the generation of a collimated laser sheet, with its centerline positioned at $y = 40$~mm. The camera is the Andor's Zyla 5.5 sCMOS,  which was equipped with a Macro Sigma lens. The lens focal length is 105~mm, and the lens aperture number was set to 2.8. Finally, a bandpass filter with a center wavelength and Full Width at Half Maximum (FWHM) of 532 and 20~nm was mounted on the camera lens. The Mie scattering field of view spans the width of the combustion chamber, and its vertical extent is limited between $y = 20$ and 60~mm. The lower extent of the Mie scattering field of view was selected to minimize reflections from the spray injector in the Mie scattering images. The effective resolution of the Mie scattering images was 28.8~$\mu$m as discussed in Appendix~B.
 
\begin{figure*}[!t]
	\centering
	\includegraphics[width = 1\textwidth]{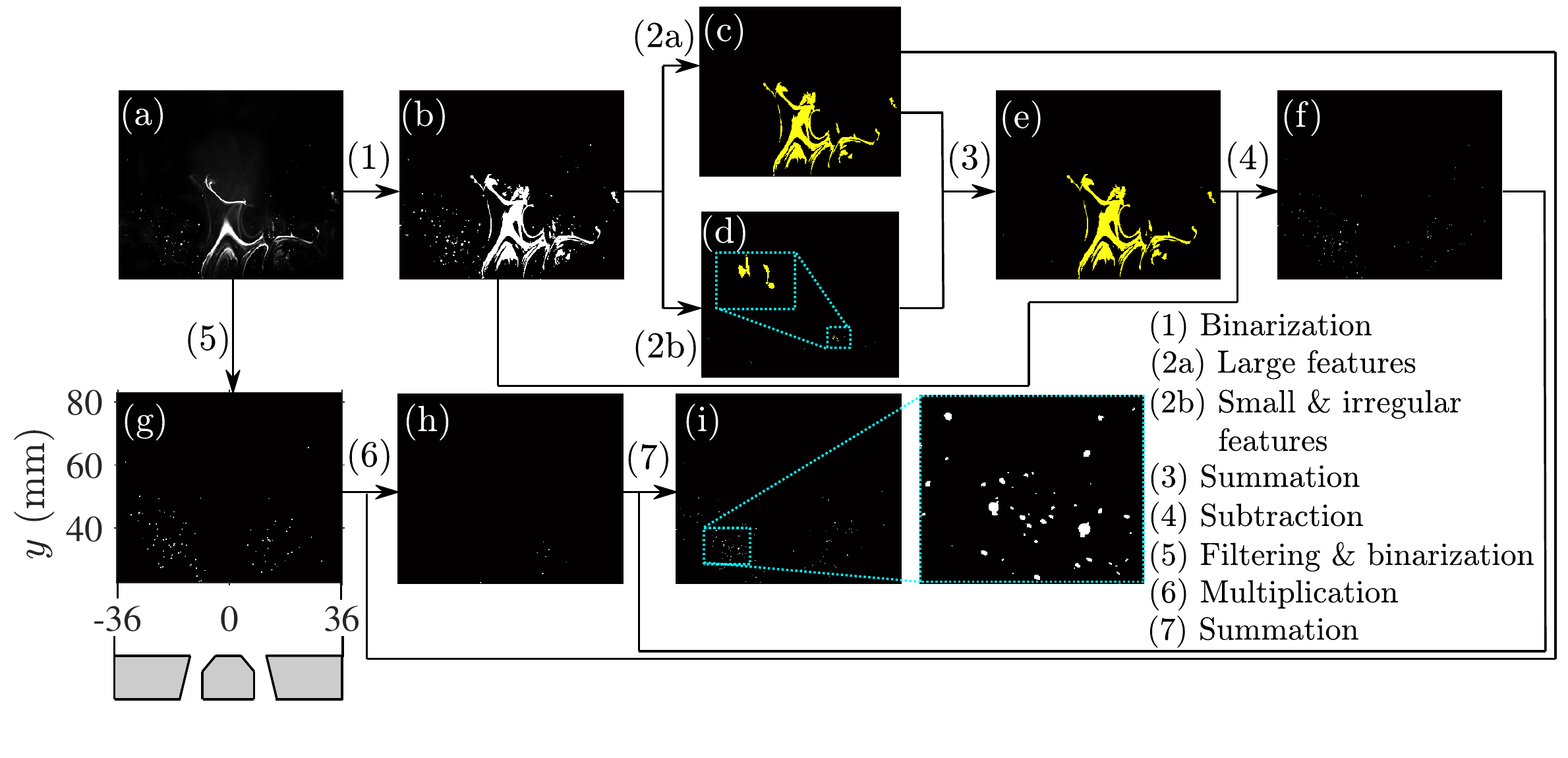}
	\caption{(a) Representative raw Mie scattering image. (b) The binarized image in (a). (c) is the large scale features in (b). (d) small scale and elongated/non-circular structures in (b). (e) is the summation of (c) and (d). (f) is (e) subtracted from (b). (g) is the mean-based filtered around the local maxima of the image shown in (a) and then binarized. (h) is the multiplication of (c) and (g) to generate droplets binarized images in the large soot structures. (i) is the summation of (h) and (f).}
	\label{fig:Miescatteringdatareduction}
\end{figure*}

The Mie scattering images were pre-processed to obtain the spray number of droplets. First, 500 images were collected and averaged (referred to as the background image) when the laser was turned off but the spray was lit. A raw Mie scattering image (corresponding to the test condition of J100M0) subtracted by the above background is shown in Fig.~\ref{fig:Miescatteringdatareduction}(a). After the background subtraction, the results were binarized (procedure 1), see Fig.~\ref{fig:Miescatteringdatareduction}(b). Analysis of the shadowgraphy images (discussed later) suggests that the spray droplets are rather spherical and relatively small. Thus, structures that are not circular and/or relatively large are not droplets and were removed from the Mie scattering data. Two separate procedures, see (2a) and (2b) in Fig.~\ref{fig:Miescatteringdatareduction}, were followed to identify large as well as small and non-spherical structures, respectively.  A labeling algorithm in MATLAB was used to identify the former type of the structures, which are shown by the yellow color in Fig.~\ref{fig:Miescatteringdatareduction}(c). As for the latter type of the structures, first, an equivalent diameter which equals the mean of the structure width and height was obtained. Then, the area of a circle with the above equivalent diameter was calculated. For structures with their shape close to a circle, the calculated area is close to the area of the structure. However, the areas of irregular (non-droplet) structures are significantly smaller than the area of the equivalent circle. For example, for an irregular structure which is 2 pixels wide and 48 pixels high, the area is $2\times48= 96~\mathrm{pixels}^2$, but the area of the equivalent circle is $(\pi/4)[(2+48)/2]^2 = 491~\mathrm{pixels}^2$, which is significantly larger than $96~\mathrm{pixels}^2$. The above criterion was used to identify the small and non-droplet structures, with a sample shown in Fig.~\ref{fig:Miescatteringdatareduction}(d). The large-scale and irregular structures were combined, procedure (3), and shown in Fig.~\ref{fig:Miescatteringdatareduction}(e). This image was then subtracted from the binarized Mie scattering image, procedure (4), with the results shown in Fig.~\ref{fig:Miescatteringdatareduction}(f). Although the results in Fig.~\ref{fig:Miescatteringdatareduction}(f) allow for identifying the majority of the droplets, the utilized algorithm automatically removes many droplets that reside inside the large-scale structures. In order to avoid loss of these droplets in the pre-processing of the Mie scattering images, they were further treated. First, the local maxima in Fig.~\ref{fig:Miescatteringdatareduction}(a) were obtained, a mean-based filtering algorithm was applied to Fig.~\ref{fig:Miescatteringdatareduction}(a) around the local maxima, and the resultant image was binarized (procedure 5), see Fig.~\ref{fig:Miescatteringdatareduction}(g). Then, the resultant image was multiplied by the mask in Fig.~\ref{fig:Miescatteringdatareduction}(c), procedure 6, to identify the small droplets inside the large-scale structures. The corresponding image is shown in Fig.~\ref{fig:Miescatteringdatareduction}(h). Finally, Fig.~\ref{fig:Miescatteringdatareduction}(h) was added to Fig.~\ref{fig:Miescatteringdatareduction}(f), see procedure 7, with the final reduced Mie scattering image of the droplets shown in Fig.~\ref{fig:Miescatteringdatareduction}(i). The inset of Fig.~\ref{fig:Miescatteringdatareduction}(i) presents a sample of the identified droplets. Procedures (1--7) were applied to all Mie scattering images and a labeling algorithm in MATLAB was used to calculate the number of the droplets, referred to as $n$, in the processed Mie scattering images.
 
\subsubsection{Shadowgraphy imaging}
\begin{figure*}[!t]
	\centering
	\includegraphics[width = 1\textwidth]{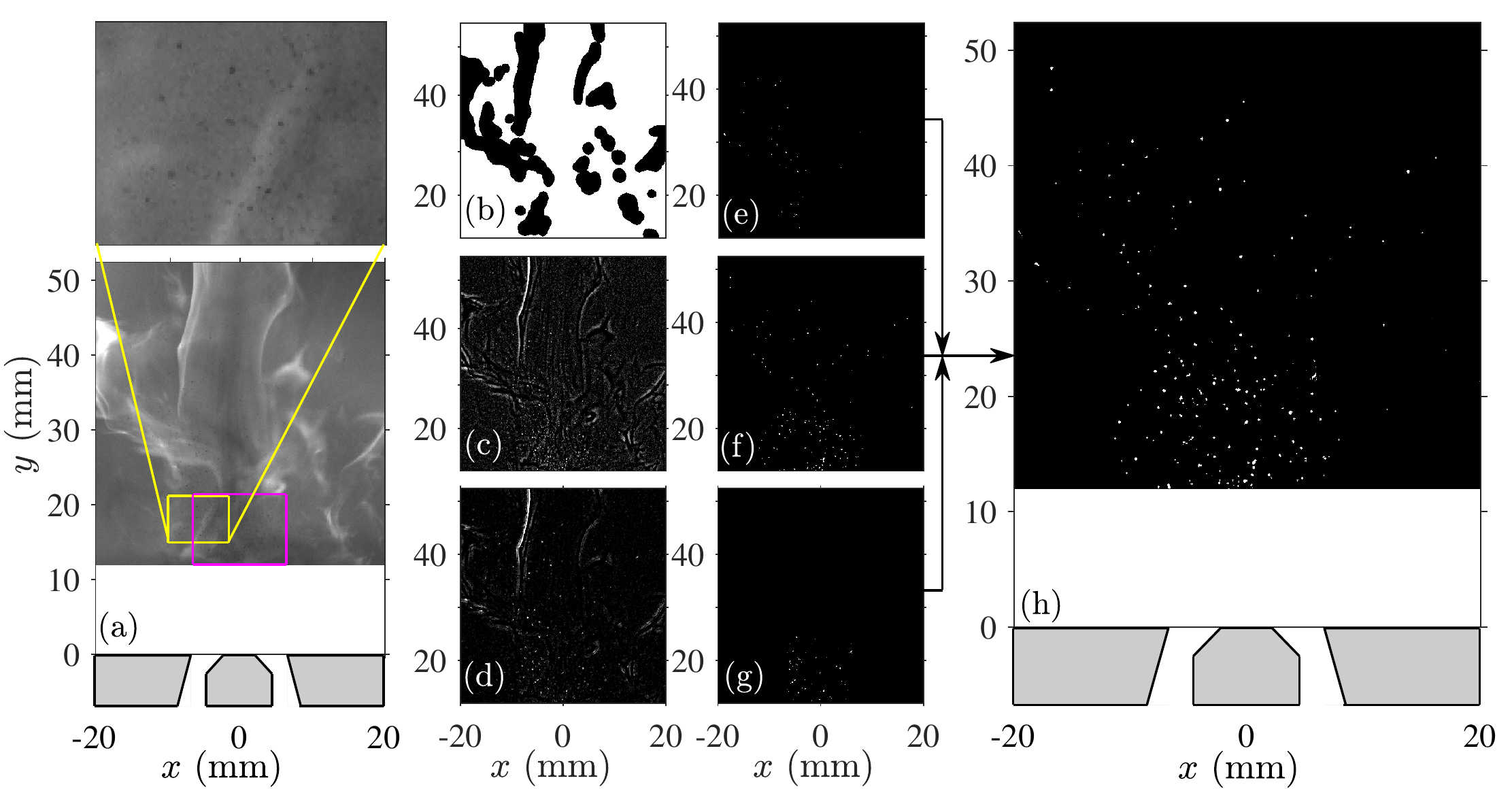}
	\caption{(a) Raw shadowgraphy image. (b) is the mask for identifying regions with large local variation of the light intensity, see the black colored region. (c) is the ratio of (a) to its mean-based filtered image. (d) is the processed image in (a) to reduce the effect of background with large intensity variation. (e) is the results in (d) filtered to exclude large as well as small non-droplet features and then multiplied by the mask in (b). (f) and (g) are the identified droplets inside the small intensity varying regions and the droplets closest to the injector, respectively. (h) is the logical sum of the results in (e--g).}
	\label{fig:shadowgraphy}
\end{figure*}

The hardware for the shadowgraphy imaging includes the high-speed camera (item 6 in Fig.~\ref{fig:diagnostics}) which is equipped with a Macro Sigma lens (with a focal length of 105~mm and aperture number of 2.8), an NL-360ARC Neewer LED lamp, and a Semrock FF01-433/530-25 dual bandpass filter. This filter was selected following the recommendations of Bennewitz \textit{et al.}~\cite{bennewitz2019systematic,bennewitz2021combustion} and is shown~\cite{mosadegh2022role} to improve the quality of the  shadowgraphy images for sooty droplets. A representative shadowgraphy image corresponding to the test condition of J100M0 is presented in Fig.~\ref{fig:shadowgraphy}(a). The dark regions in the inset of the figure are the shadow of the droplets. The pixel size for the shadowgraphy experiments was 68.0~$\mu$m. The effective spatial resolution was determined using the USAF~1951 target plate and was 99.2~$\mu$m (see Appendix~B). In the present study, two challenges exist for identifying the droplets using the shadowgraphy technique. First, due to the presence of the turbulent flames and soot, the detected light intensity can significantly vary in the spray region. Specifically, some regions near the droplets may feature a relatively small intensity gradient; however, some regions that entail the ridges of the flame and/or soot structures may feature relatively large local intensity gradients. Both of these posed challenges for identifying the droplets, with the corresponding regions were treated using separate strategies. The second challenge is that the spray is relatively dense near the injector, which leads to the overlapping of the droplets shadows. 


To help addressing the first challenge, regions with large variations in the background light intensity were first identified by calculating the absolute value of the difference between the pixel intensity and the local mean. Then, a mean-based filter and a threshold were applied to the resultant image, creating a mask shown in Fig.~\ref{fig:shadowgraphy}(b). To identify the droplets in the large-intensity-variation regions (see the black color in Fig.~\ref{fig:shadowgraphy}(b)), the absolute value of the raw image gradient was obtained. Then, the ratio of the absolute intensity gradient and the absolute mean local gradient were calculated. This was followed by applying a mean-based filter to the above ratio. This ratio was multiplied by the ratio of the raw image to its mean-based filter, with the latter ratio and the resultant image shown in Figs.~\ref{fig:shadowgraphy}(c) and (d), respectively. The results in Fig.~\ref{fig:shadowgraphy}(d) were then filtered by that developed in section~\ref{subsubsection:mie} (see procedures (2a) and (2b) in Fig.~\ref{fig:Miescatteringdatareduction}) to identify and remove large aspect ratio structures, which are not droplets. The resultant image includes droplets in both small and large-intensity-variation regions. Since the detection of the droplets in the large-intensity-variation regions are of interest in this step, the resultant image was then multiplied by the mask in Fig.~\ref{fig:shadowgraphy}(b), with black being unity and white being zero. The product is shown in Fig.~\ref{fig:shadowgraphy}(e).

In the above procedure, a relatively large threshold is applied to Fig.~\ref{fig:shadowgraphy}(d) in order to avoid false detection of large and circular soot structures. As a result of this, those identified in the above procedure do not include majority of the droplets. Indeed, many of the droplets reside in the white region of the mask shown in Fig.~\ref{fig:shadowgraphy}(b). In order to identify these droplets, the ratio of the pixel intensity in the raw image to that of the local mean (which was calculated in the previous step and shown in Fig.~\ref{fig:shadowgraphy}(c)) was utilized. The resulting image was thresholded and multiplied by the mask shown in Fig.~\ref{fig:shadowgraphy}(b), with black being zero and white being unity. The product is shown in Fig.~\ref{fig:shadowgraphy}(f).

A rectangular region, which is highlighted by the pink lines in Fig.~\ref{fig:shadowgraphy}(a) was considered to identify the droplets at the vicinity of the nozzle. Droplets within this region were identified by, first, subtracting the pixel intensities in Fig.~\ref{fig:shadowgraphy}(d) from their local mean values; and, then thresholding the resultant image. To ensure the overlapping droplets are not removed in the above procedure, the resultant image was not filtered by the aspect ratio filters (which were developed and discussed in processes (2a and 2b) in section~\ref{subsubsection:mie}. Finally, the identified droplets within the rectangular pink mask highlighted in Fig.~\ref{fig:shadowgraphy}(a) were considered, with the resultant image shown in Fig.~\ref{fig:shadowgraphy}(g). The results in Figs.~\ref{fig:shadowgraphy}(e), (f), and (g) were added and repeating droplets were removed (to avoid double counting) and the final image is shown in  Fig.~\ref{fig:shadowgraphy}(h). This is the final processed shadowgraphy image from the raw image (shown in Fig.~\ref{fig:shadowgraphy}(a)) that contains all identified droplets. Using the post-processed shadowgraphy images, the number of droplets ($n_\mathrm{S}$) as well as individual droplet diameter $d_\mathrm{S} = \sqrt{4A_\mathrm{S}/\pi}$, with $A_\mathrm{S}$ being individual droplet area, were obtained.

\subsubsection{Interferometric Laser Imaging for Droplet Sizing}
Separate Interferometric Laser Imaging for Droplet Sizing was employed in the present study to characterize the spray. This technique has been used for measuring the droplet diameter for both reacting and non-reacting flows in the past \cite{qieni2016high,bocanegra2015measuring,renoux2018experimental}. In this technique, the interference of reflected and first-order refracted rays scattered from a spherical droplet illuminated by a laser source is used to measure the droplet diameter. The utilized hardware is identical to those used for the Mie scattering, except, a high-precision rotational stage (see item~15 in Fig.~\ref{fig:diagnostics}) was also used to adjust the angular position of the camera with respect to the laser sheet.



 
The collected ILIDS images are analyzed to obtain the spray droplet diameters using \cite{qieni2016high,bocanegra2015measuring,renoux2018experimental}.

 \begin{equation}
 \label{eq:ILIDS}
 	d=\dfrac{2\lambda N}{\alpha}\left[\cos(\frac{\theta}{2})+\dfrac{m\sin(\theta/2)}{\sqrt{m^{2}-2m\cos(\theta/2)+1}}\right]^{-1}.
 \end{equation}
In Eq.~(\ref{eq:ILIDS}), $\lambda $ is the wavelength of the incident laser light, which is 532~nm. $m$ is Jet A-1 index of refraction, which is taken from \cite{verma2018detection} and is 1.44. $\theta$ is the angle between the normal to the camera lens and the laser sheet. This angle was set to $70^\mathrm{o}$ to maximize the quality of the collected fringe patterns, following the recommendations of Sahu \textit{et al.}~\cite{sahu2016droplet}. The collection angle, $\alpha$, depends on the lens diameter ($d_\mathrm{l}$, which is 60~mm) and the working distance ($L$, which is the distance between the plane of the camera lens and the laser sheet). Specifically, this angle is calculated using $\alpha=2 \arctan [d_\mathrm{l}/(2L)]$. In the present study, $L = 240$~mm is fixed, and as a result, the collection angle is $14.3^{\mathrm{o}}$ ($0.25~\mathrm{rad}$). In Eq.~(\ref{eq:ILIDS}), $N$ is the number of the fringe patterns, which is obtained using the following data reduction procedure.

\begin{figure*}[!t]
	\centering
	\includegraphics[width = 1\textwidth]{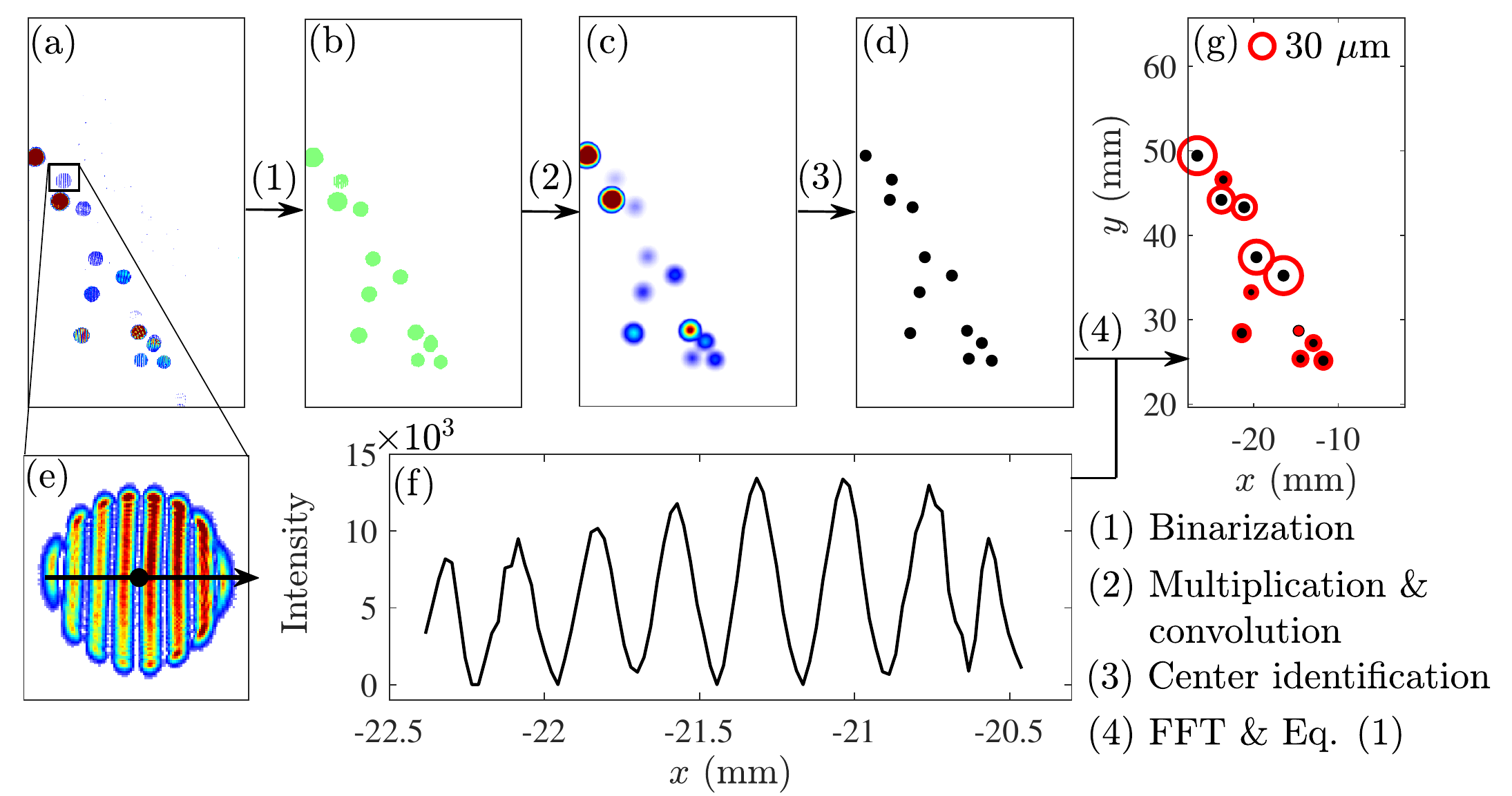}
	\caption{(a) Representative raw ILIDS image. (b) Binarized image corresponding to (a). (c) Convoluted image of the results in (a) multiplied by (b). (d) Location of the droplets centers. (e) and (f) are sample 2D light intensity variation and that along the direction normal to the fringe patterns, respectively. (g) Droplets centers and their corresponding size (the red circle diameter highlighted by $30~\mathrm{\mu}$m is to scale the diameter of the droplets identified in the figure).}
	\label{fig:datareductionILIDS}
\end{figure*}

First, while the laser was turned off, 500 background images were collected and subtracted from the raw ILIDS images. Figure.~\ref{fig:datareductionILIDS}(a) presents a representative ILIDS image subtracted by the background. Then, the raw image was binarized, see Fig.~\ref{fig:datareductionILIDS}(b). The background subtracted image (Fig.~\ref{fig:datareductionILIDS}(a)) was multiplied by the binarized image in Fig.~\ref{fig:datareductionILIDS}(b). Then, a disk-shaped convolution function in MATLAB was applied to the obtained results, which allowed to maximize the intensity at the center of the droplets, similar to the procedure used in~\cite{bocanegra2015measuring}. The obtained convoluted image is shown in Fig.~\ref{fig:datareductionILIDS}(c). The centers of the droplets were then obtained, with the corresponding results shown in Fig.~\ref{fig:datareductionILIDS}(d). A representative droplet with its identified center is shown in Fig.~\ref{fig:datareductionILIDS}(e). Variation of the light intensity along the direction normal to the fringe patterns (shown in Fig.~\ref{fig:datareductionILIDS}(f)) was obtained and the Fast Fourier Transform of the intensity was used to identify the number of the fringe patterns. Figure~\ref{fig:datareductionILIDS}(g) presents the centers of the identified droplets in Fig.~\ref{fig:datareductionILIDS}(a) as well as the overlaid red circles with diameters scaled to reflect the size of the identified droplets. As discussed in Appendix~B, the effective resolution of the ILIDS images was 21.4~$\mu$m, which allowed for detecting droplets with diameters $5 \lesssim d \lesssim 107~\mathrm{\mu}$m.

\subsection{Test Conditions}
Three conditions were tested and tabulated in Table~\ref{Tab:Testedconditions}. For all test conditions, both the total generated power and the global fuel-air equivalence ratio were kept constant at 10~kW and 0.6, respectively. In the table, J0M100 and J100M0 pertain to test conditions for which the fuel was either methane or Jet A-1, respectively. For J40M60, the mass flow rates of methane and Jet A-1 were set to produce 40$\%$ (4~kW) and 60$\%$ (6~kW) of the total power from Jet A-1 and methane, respectively. In the above calculation, the lower heating values of methane and Jet A-1 were used and set to 49853 and 43200 kJ/kg, respectively. The set pressure inside the liquid fuel reservoir ($P_\mathrm{l}$) and the fuels ($\dot{m}_\mathrm{CH4}$ and $\dot{m}_\mathrm{Jet~ A-1}$) and air ($\dot{m}_\mathrm{air}$) mass flow rates are provided in the second to fifth columns of the table. Several combinations of the gaseous and liquid fuels flow rates were considered for measurements, however, the combination of powers generated by gas and liquid fuels for the test condition of J40M60 was selected since the corresponding flow rate of Jet A-1 (5.6 grams per minute) was the minimum liquid flow rate required to form the spray.

\begin{table}[!htbp]
	\caption{Test conditions. The total power (10~kW) and the global fuel-air equivalence ratio (0.6) were fixed for all test conditions. The units of $P_\mathrm{l}$ and $p^\prime_\mathrm{rms}$ are kPa and Pa, respectively. The unit of $\dot{m}_\mathrm{CH4}$, $\dot{m}_\mathrm{Jet~A-1}$, and $\dot{m}_\mathrm{air}$ is grams per minute.}
	\label{Tab:Testedconditions}
	\centering
	\scalebox{1.0}{
		\begin{tabular}{cccccc}
			\hline
			\hline
			Condition & $P_\mathrm{l}$ & $\dot{m}_\mathrm{CH4}$ & $\dot{m}_\mathrm{Jet~A-1}$ & $\dot{m}_\mathrm{air}$ & $p'_\mathrm{rms}$ \\
            \hline

            J0M100 & 0 & 12.0 & 0 & 343.4 & 28.5 \\
            \hline
            
            J40M60 & 18.4 & 7.2 & 5.6 & 340.9 & 46.3 \\
            \hline

            J100M0 & 107.1 & 0 & 13.9 & 337.0 & 83.5 \\

			\hline
			\hline
	\end{tabular}}
\end{table}

\section{Pressure, flame chemiluminescence, and spray characteristics}
\label{characterization}
\begin{figure*}[!t]
	\centering
\includegraphics[width = 0.5\textwidth]{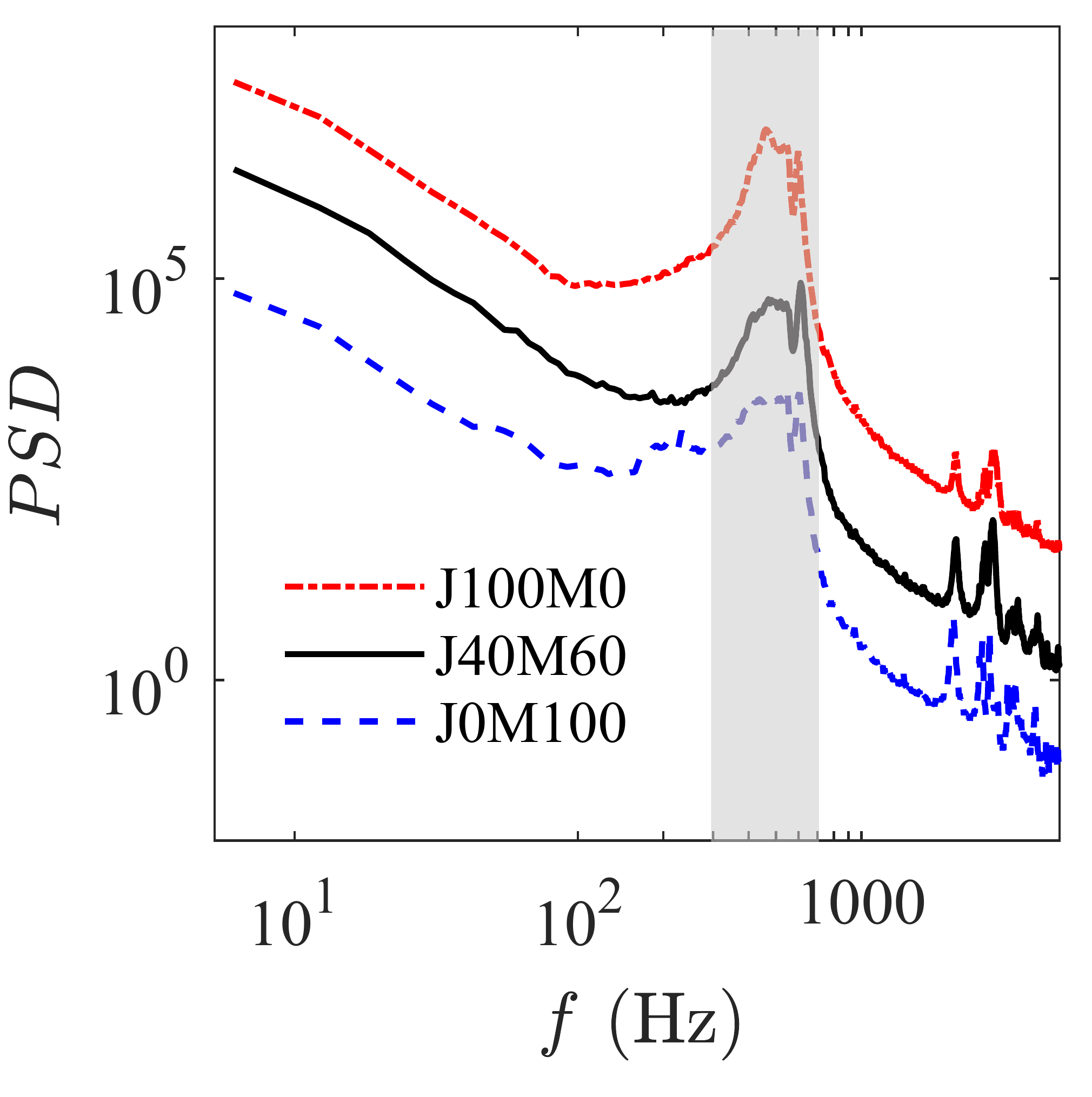}
	\caption{The power spectrum density of the plenum pressure oscillations. The results are stepped by a factor of 10 for clarity.}
	\label{fig:pressure}
\end{figure*}

Addressing the objective of the present study requires understanding the characteristics of the plenum pressure, flame chemiluminescence, and spray. These characteristics are studied in this section. As presented in the last column of Table~\ref{Tab:Testedconditions}, changing the test condition from J0M100 to J40M60 and J100M0 increases the root-mean-square (RMS) of the pressure fluctuations from 28.5 to 46.3 and 83.5, respectively. Figure~\ref{fig:pressure} presents the power spectrum densities ($PSD$) of the plenum pressure fluctuations for all test conditions. The results in the figure are stepped by a factor of 10 for clarity. As can be seen, the plenum pressure features large amplitude oscillations near $300 \lesssim f \lesssim 700$~Hz for all test conditions. This frequency band is highlighted by the gray shaded area in Fig.~\ref{fig:pressure}.

\begin{figure*}[!t]
	\centering
	\includegraphics[width = 1\textwidth]{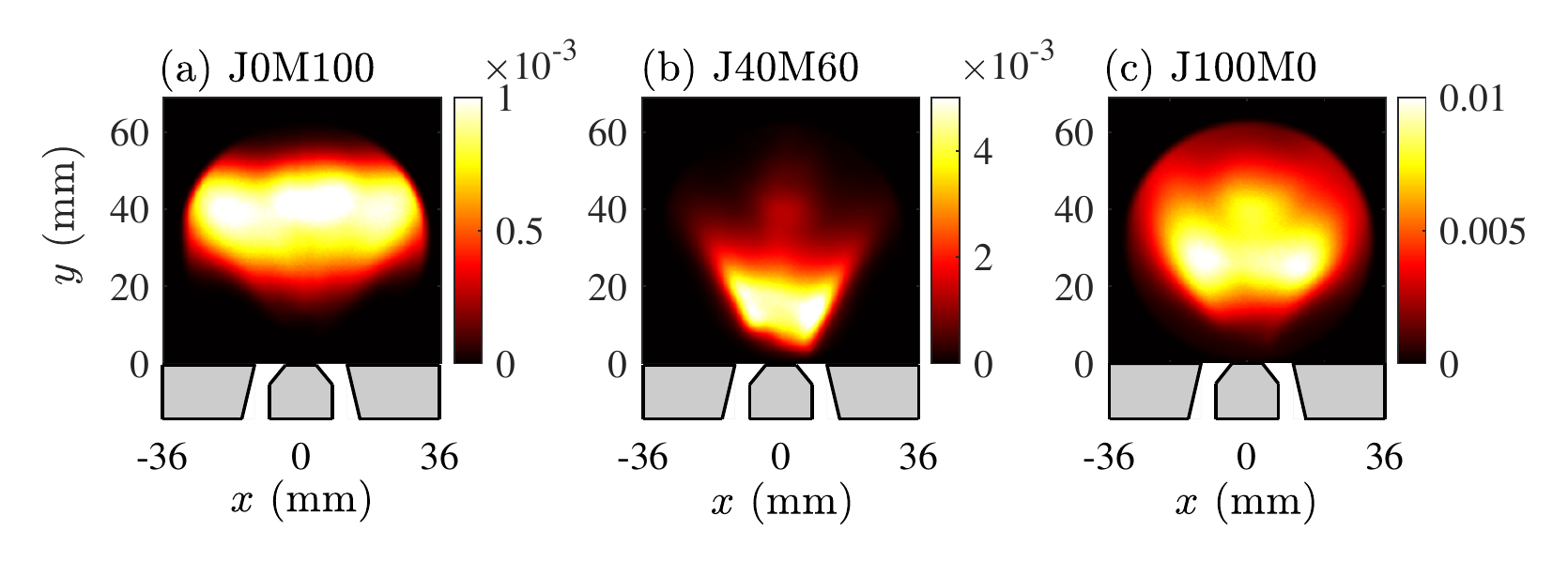}
	\caption{Time-averaged flame chemiluminescence for (a) J0M100, (b) J40M60, and (c) J100M0.}
	\label{fig:OHspatialmean}
\end{figure*}

The time averaged flame chemiluminescence ($\overline{CL}$) for the test conditions J0M100, J40M60, and J100M0 are presented in Figs.~\ref{fig:OHspatialmean}(a--c), respectively. For fully premixed flames (J0M100), the maximum mean flame chemiluminescence is positioned at $y \approx 40$~mm. Adding the Jet A-1 spray and decreasing the methane flow rate displaces the mean flame chemiluminescence closer to the nozzle exit plane. Also, the maximum $\overline{CL}$ increases by a factor of about 5 and 10 changing the test condition from J0M100 to J40M60 and J100M0, respectively. The broadband luminosity images were also collected (not shown here), and it was observed that the spray flames of test conditions J100M0 and J40M60 feature large soot emissions, especially close to the spray injector. Thus, the large values of $\overline{CL}$ for J40M60 and J100M0 (compared to J0M100) as well as the smaller vertical position of the maximum $\overline{CL}$ may be due to the pronounced soot formation close to the injector. Nonetheless, for the analyses presented here, the flame chemiluminescence is not used as a quantitative marker of the heat release rate. Instead, the flame chemiluminescence is used for the purposes of training and predicting the number of spray droplets data.

\begin{figure*}[!t]
	\centering
	\includegraphics[width = 0.75\textwidth]{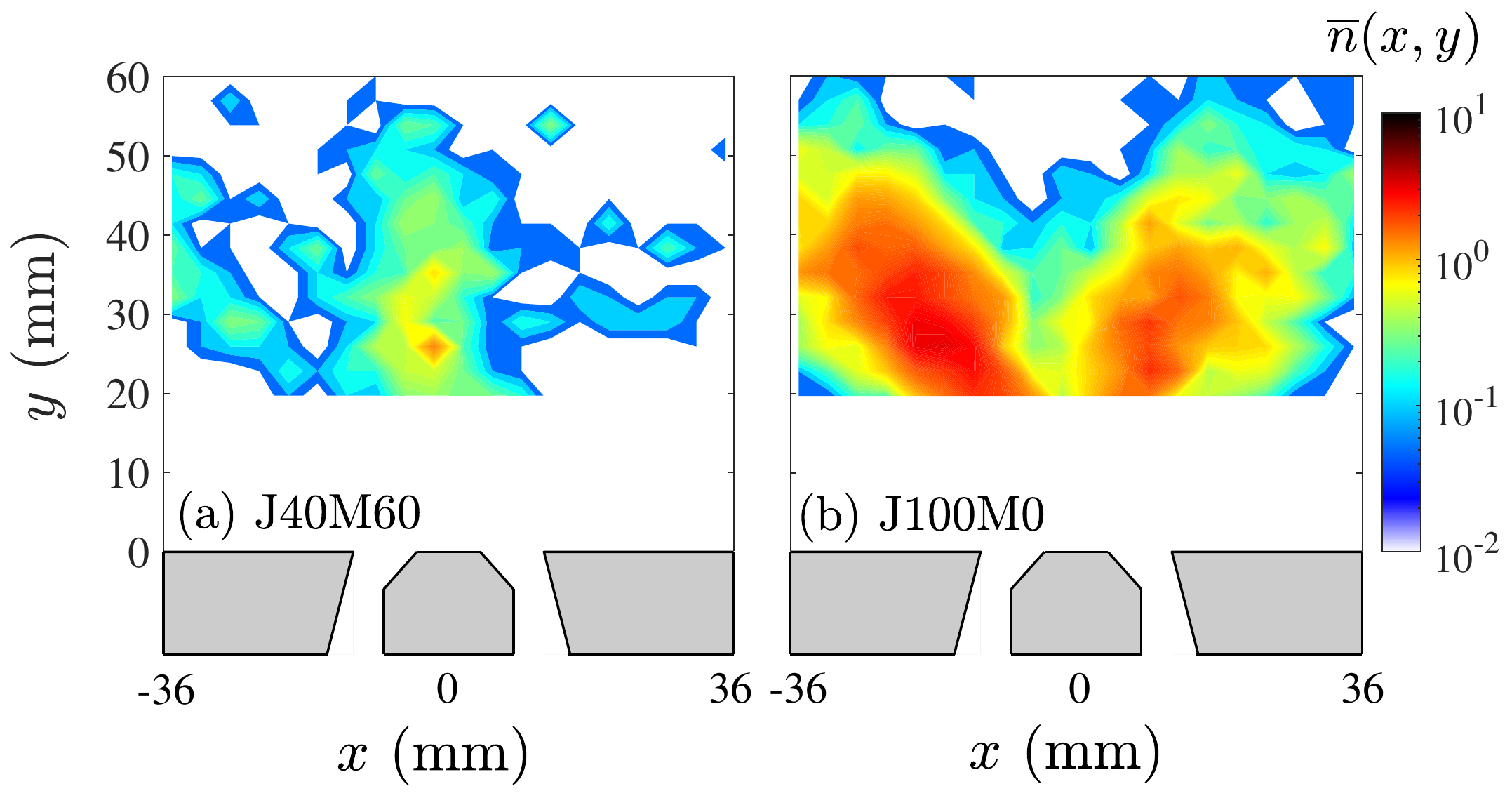}
	\caption{The mean number of droplets, $\overline{n}(x,y)$ obtained from the Mie scattering measurements. (a) and (b) correspond to test conditions J40M60 and J100M0, respectively.}
	\label{fig:Miend}
\end{figure*}

In the present study, the number of droplets inside a laser-illuminated plane ($n$) is studied using the Mie scattering technique. Additionally, using the shadowgraphy technique, the number and diameter of the droplets inside a focusing volume are investigated. As discussed in the following, both techniques are necessary and provide complementary understanding of the spray characteristics and/or dynamics. Analysis of the Mie scattering images suggests that the mean (standard deviation) of the number of droplets calculated for all frames and for the test conditions J40M60 and J100M0 are 28 (23) and 204 (45), respectively. Figures~\ref{fig:Miend}(a) and (b) present the mean of the number of droplets obtained from the Mie scattering images for the test conditions J40M60 and J100M0, respectively. The contours are presented in a logarithmic scale for improved clarity of presentation. As can be seen, for J40M60 and J100M0, the spray extends to about 60~mm downstream of the nozzle. J100M0 features a conical spray; however, for J40M60, the spray is mostly injected along the centerline. Although the spray for the test condition of J40M60 features a non-conical shape, the number of droplets can be estimated for this condition and the shape of the spray does not negatively impact such estimation. 

\begin{figure*}[!t]
	\centering
	\includegraphics[width =0.9 \textwidth]{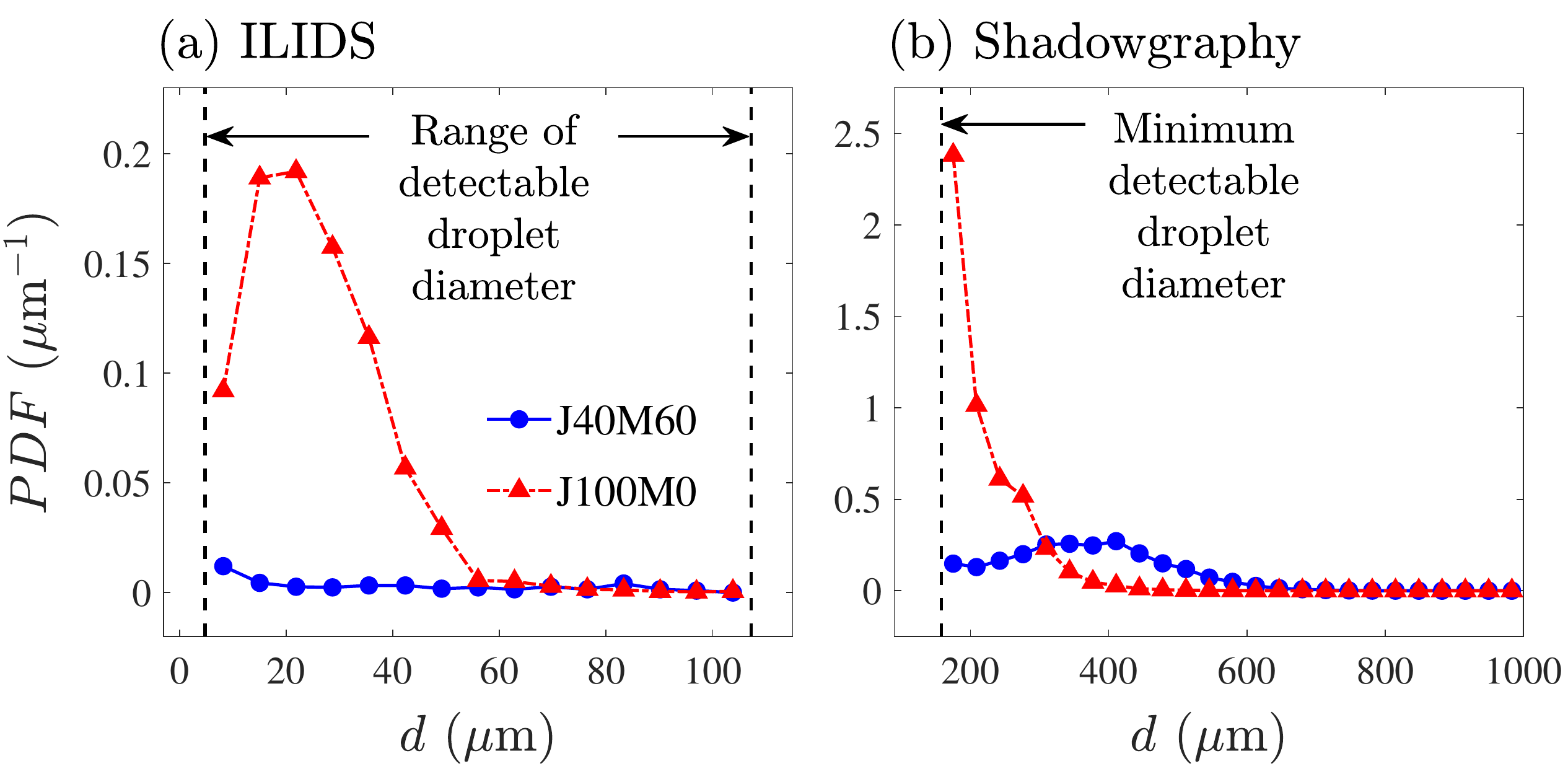}
	\caption{The probability density functions of the droplet diameter estimated using (a) the ILIDS and (b) the shadowgraphy techniques.}
	\label{fig:ILIDSsize}
\end{figure*}

The Probability Density Function (PDF) of the droplets diameters obtained from the ILIDS and shadowgraphy techniques are presented in Figs.~\ref{fig:ILIDSsize}(a)~and~(b), respectively. Generally, the values of the PDF reported in Fig.~\ref{fig:ILIDSsize}(a) are smaller than those in Fig.~\ref{fig:ILIDSsize}(b). This is because the shadowgraphy technique allows for detecting the droplets inside a relatively large volume (compared to ILIDS); and as a result, a large number of droplets (hence larger PDF values) are reported in Fig.~\ref{fig:ILIDSsize}(b) compared to those in Fig.~\ref{fig:ILIDSsize}(a). As discussed in section~\ref{EM} as well as in Appendix~B, the range of the detectable droplets diameter for ILIDS is $5~\mathrm{\mu m} \lesssim d \lesssim 107~\mathrm{\mu m}$. For a droplet to be detected by the shadowgraphy technique, its area should be at least twice the area of an effective pixel resolution. Thus, the minimum detectable droplet area is $2\times 99.2^2~(\mu \mathrm{m})^2= 19681~(\mu\mathrm{m})^2$. This leads to a minimum detectable diameter of $\sqrt{4A_\mathrm{min}/\pi}=158~\mathrm{\mu m}$ for the shadowgraphy technique. With the exception of $107~\mathrm{\mu m} \lesssim d \lesssim 158~\mathrm{\mu m}$, and for their corresponding detectable droplet diameter range, the combination of the ILIDS and the shadowgraphy techniques allows for understanding the distribution of the droplet diameters in the present study. The PDF of the droplet diameter for the test conditions J40M60 and J100M0 are shown by the blue circular and red triangular data symbols, respectively, in Fig.~\ref{fig:ILIDSsize}. For droplets smaller than 107~$\mu$m, the most probable droplet diameter is about 22~$\mu$m for J100M0. Since the injection pressure for the test condition of J40M60 is about five times smaller than that for J100M0 (see Appendix~A), and as a result, the total number of the generated droplets averaged over all collected Mie scattering images for J40M60 is significantly smaller than that for J100M0 (150 versus 2902), the PDF does not allow for discerning a most probable diameter for J40M60 and for $5~\mathrm{\mu m} \lesssim d \lesssim 107~\mathrm{\mu m}$. The results in Fig.~\ref{fig:ILIDSsize}(b) suggest that the PDF of the droplet diameter decreases with increasing the diameter for the test condition of J100M0. However, for $158~\mathrm{\mu m} \lesssim d \lesssim 1000~\mathrm{\mu m}$, J40M60 features a most probable droplet diameter of about 411~$\mathrm{\mu m}$. In essence, the results presented in Fig.~\ref{fig:ILIDSsize} suggest that changing the test condition from J100M0 to J40M60 changes the most probable spray droplets diameter from about 22 to 411~$\mathrm{\mu m}$.

\begin{figure*}[!t]
	\centering
	\includegraphics[width = 1.0 \textwidth]{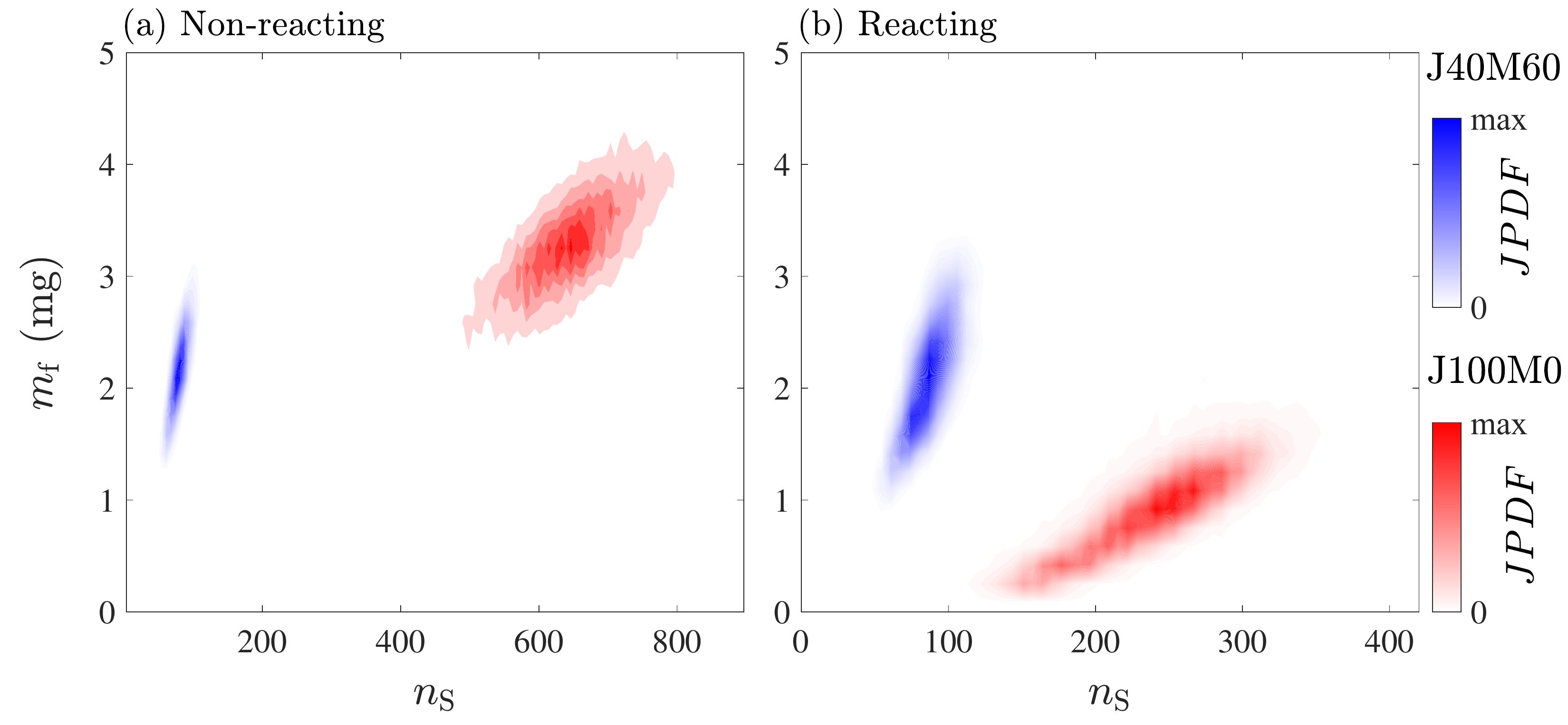}
	\caption{The joint probability density function of the fuel mass ($m_\mathrm{f}$) versus the number of droplets ($n_\mathrm{S}$), both estimated based on the shadowgraphy technique. (a) and (b) correspond to the non-reacting and reacting conditions, respectively.}
	\label{fig:mvsn}
\end{figure*}

Although the probability density functions generated from the ILIDS and shadowgraphy techniques provide complementary information related to the droplet characteristics, the temporal variation of such characteristics can only be inferred from the shadowgraphy technique in the present study, as the utilized laser features a maximum repetition rate of 10~Hz, which is relatively small. Acknowledging the minimum detectable droplet diameter from the shadowgraphy technique (158~$\mu$m) and using the measured density of Jet A-1 at the laboratory temperature ($\rho_\mathrm{l}=812.0~\mathrm{kg/m^3}$), the total mass of the droplets within the focusing volume of the shadowgraphy technique, $m_\mathrm{f}$, can be estimated using

\begin{equation}
	\label{eq:mf}
	m_\mathrm{f}(t) = \rho_\mathrm{l}\sum_{i = 1}^{i = n_\mathrm{S}(t)} \frac{\pi}{6} d^3(i),
\end{equation}
where $n_\mathrm{S}(t)$ is the total detected number of droplets at time $t$ using the shadowgraphy technique. Figures~\ref{fig:mvsn}(a) and (b) present the Joint Probability Density Functions (JPDF) of the total detected droplets mass versus the total number of droplets for non-reacting and reacting conditions, respectively. The results suggest that the mean mass of the resolved droplets for the test condition J40M60 decreases from 2.2 to 2.1~mg changing from non-reacting to reacting conditions; and, $m_\mathrm{f}$ decreases from 3.3 to 0.9~mg changing from non-reacting to reacting conditions for J100M0. In addition to the droplets evaporation (due to combustion), the above decrease is due to the change in the shape of the spray after it is lit. Nonetheless, The results pertaining to both non-reacting and reacting conditions suggest that there exists a positive correlation between $m_\mathrm{f}$ and the number of droplets ($n_\mathrm{S}$), both obtained from the shadowgraphy technique for droplets larger than 158~$\mu$m. Given the above positive correlation, it is speculated that provided a dominant instability exists in the temporal variation of $n_\mathrm{S}$, such instability may also be present in the temporal variation of the mass of the liquid fuel inside the combustor.

\begin{figure*}[!t]
	\centering
	\includegraphics[width = 1\textwidth]{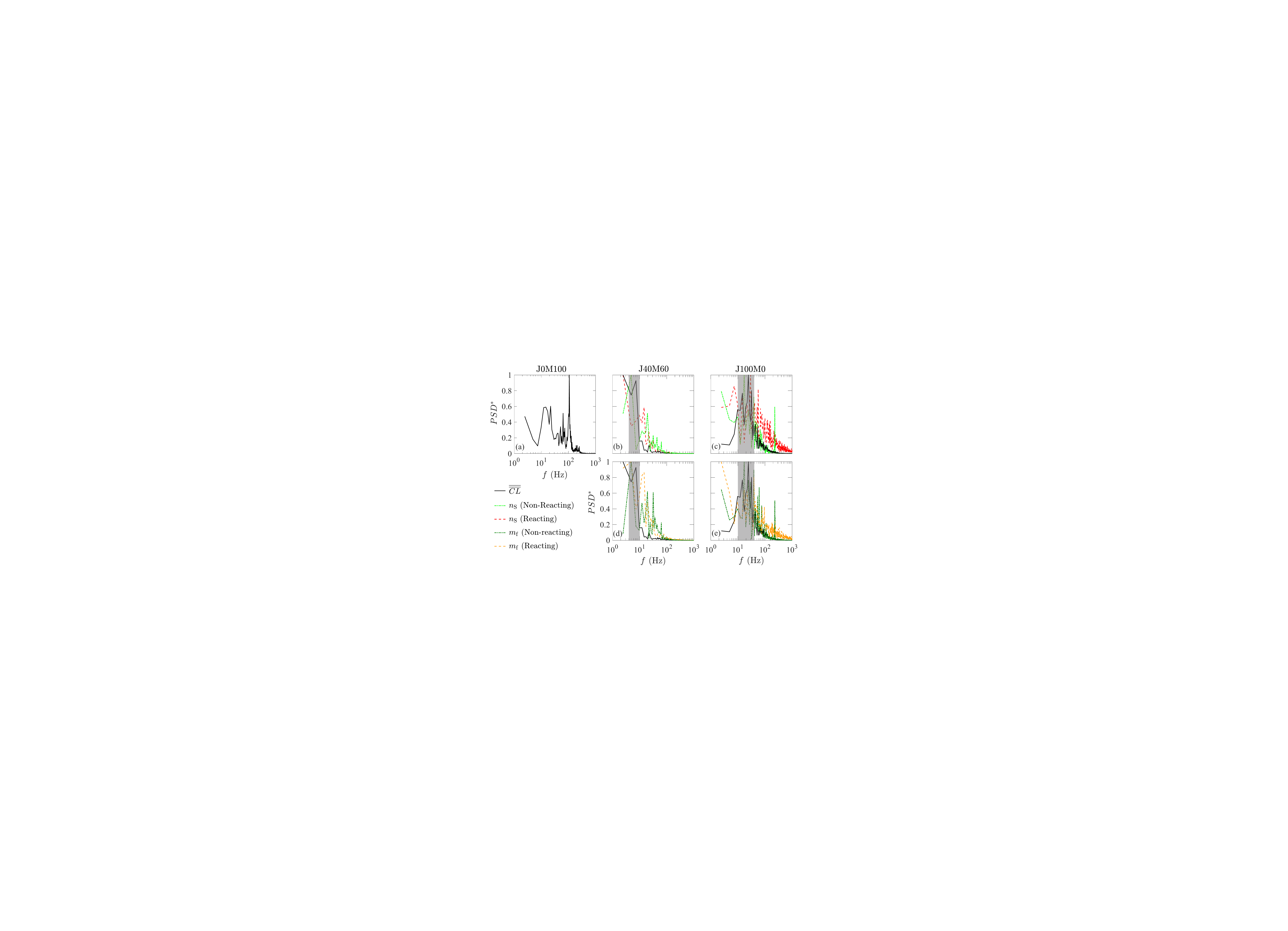}
	\caption{The normalized power spectrum densities of the spatially averaged flame chemiluminescence (solid black lines), the number of droplets for non-reacting spray (dotted-dashed light green lines), the number of droplets for reacting spray (dotted-dashed red lines), the mass of the droplets for non-reacting spray (dotted-dashed dark green lines), and the mass of the droplets for reacting spray (dashed orange lines). The results in (a), (b and d), and (c and e) correspond to test conditions J0M100, J40M60, and J100M0, respectively.}
	\label{fig:PSDs}
\end{figure*}

In order to assess the above speculation, the power spectrum densities of the number, $PSD(n_\mathrm{S})$, and mass, $PSD(m_\mathrm{f})$, of the droplets normalized by the corresponding maxima were calculated. Additionally and for comparison purposes, the normalized power spectrum density of the spatially averaged flame chemiluminescence was estimated and is presented by the solid black lines in Fig.~\ref{fig:PSDs}. The results in Figs.~\ref{fig:PSDs}(a), (b and d), and (c and e) correspond to the test conditions J0M100, J40M60, and J100M0, respectively. In the figures, the $PSD^*$ of $n_\mathrm{S}$ and for non-reacting and reacting conditions are shown by the light green dashed and red dashed lines, respectively, for both spray test conditions. The $PSD^*$ of $m_\mathrm{f}$ and for non-reacting and reacting conditions are shown by the dark dotted-dashed green and orange dashed lines, respectively. The results in Fig.~\ref{fig:PSDs} show that the flame chemiluminescence spectra features a dominant peak at $f \approx 100$~Hz for perfectly premixed methane-air flames; however, adding the spray and reducing the methane flow rate removes this dominant frequency in the power spectrum densities. For the non-reacting and reacting sprays, both $n_\mathrm{S}$ and $m_\mathrm{f}$ feature large amplitude oscillations for $f \lesssim 10$~Hz and $f\approx 10-40$~Hz corresponding to J40M60 and J100M0, respectively. Similar to the normalized power spectrum densities of the number of droplets and mass of the droplets, the flame chemiluminescence $PSD^*$ also features relatively large amplitude oscillations near the above frequencies for the corresponding spray-relevant test conditions, see the gray regions in the figure.

Comparison of the results presented in Figs.~\ref{fig:pressure}~and~\ref{fig:PSDs} show that, for the fully premixed condition (J0M100), the pressure and flame chemiluminescence feature dominant frequencies that do not match. Similar to the fully-premixed flames, for the spray test conditions, the pressure spectra do not follow the PSDs of the flame chemiluminescence, spray number of droplets, and droplets mass. However, for the spray test conditions, the chemiluminescence signal features strong oscillations at frequencies that match those of the number of droplets and the droplets mass obtained from the shadowgraphy technique for both non-reacting and reacting conditions. In essence, acknowledging the limitations of the $\mathrm{OH^*}$ chemiluminescence and shadowgraphy techniques, the flame chemiluminescence features an instability that also exists in the fuel number of droplets and their mass for both non-reacting and reacting sprays. This suggests that the flame chemiluminescence oscillations are driven by the fuel injection and that there exists a coupling between the amount of the spray in the combustor and the flame chemiluminescence.

The above analysis was performed using the shadowgraphy technique, which suffers from low imaging resolution (about 99~$\mu$m) but provides temporally resolved information regarding the number of droplets and their mass in the measurement volume. Compared to this technique, the Mie scattering technique features an improved spatial resolution (about 29~$\mu$m), which allows for estimating the number of droplets with a wide range of diameters inside the plane of measurements. The Mie scattering technique used in this study, however, suffers from low temporal resolution, since the utilized laser featured a maximum frequency of 10~Hz. In the following section, we will utilize the findings discussed above to develop a framework, which will be used in section~\ref{Results} to predict the time-resolved variation of the number of droplets from the Mie scattering technique. 




\section{The data-driven framework}
\label{framework}
\begin{figure*}[!t]
	\centering
	\includegraphics[width = 0.75\textwidth]{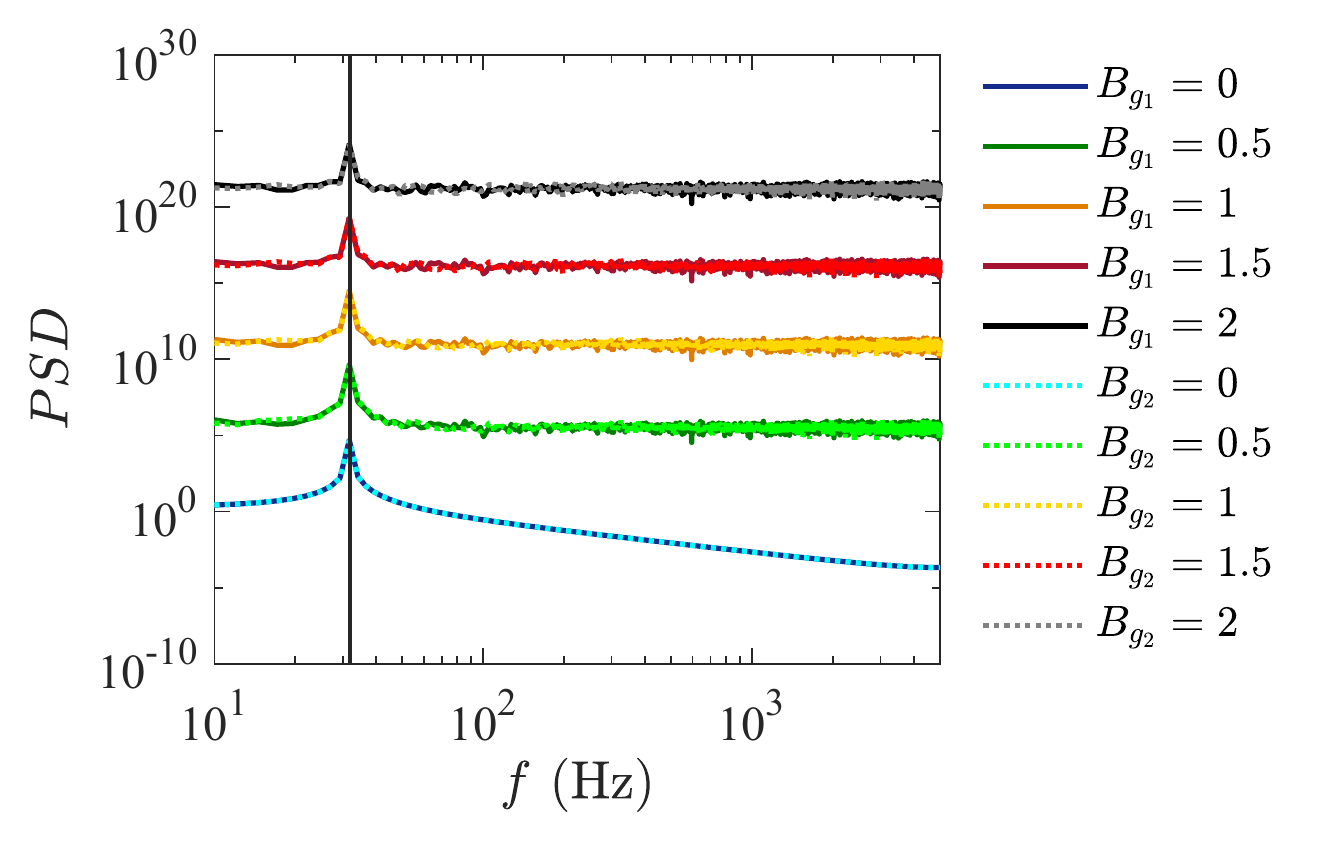}
	\caption{The power spectrum densities of $g_1$ (solid curves) and $g_2$ (dotted curves) for several background values of the random fluctuations.}
	\label{fig:PSDg1g2}
\end{figure*}

In this section, a framework is developed to predict the temporally resolved spray number of droplets from time-resolved chemiluminescence and sparse Mie scattering data. The framework is developed for two generic signals, $g_1(t)$ and $g_2(t)$, which correspond to the time-resolved and spatially averaged flame chemiluminescence as well as the number of droplets (obtained from the sparse Mie scattering measurements). In section~\ref{characterization}, it was discussed that the mass and the number of the droplets obtained from the shadowgraphy technique feature large amplitude oscillations at frequencies matching those of the chemiluminescence oscillations. In this section, it is assumed such characteristic can be extended to the number of droplets identified by the Mie scattering technique. Specifically, it is assumed that $g_1(t)$ and $g_2(t)$ feature large amplitude oscillations at a matching frequency. The validity of this assumption for prediction of $g_2(t)$ is later assessed in section~\ref{Results}. $g_1(t)$ and $g_2(t)$ were formulated as

\begin{subequations}
	\begin{equation}
		\label{eq:g1}
		g_1(t) = \sin(2\pi f t)+B_{g_1} \mathcal{R}_1(t),
	\end{equation}
	\begin{equation}
		\label{eq:g2}
		g_2(t) = \sin(2\pi f t)+B_{g_2} \mathcal{R}_2(t).
	\end{equation}
\end{subequations}
In Eqs.~(\ref{eq:g1})~and~(\ref{eq:g2}), $f$ is the matching frequency at which both $g_1(t)$ and $g_2(t)$ feature relatively large amplitude oscillations. This frequency was set to 31.8~Hz and corresponds to the test condition of J100M0. In Eqs.~(\ref{eq:g1})~and~(\ref{eq:g2}), $\mathcal{R}_1$ and $\mathcal{R}_2$ are random functions that vary between -1 and 1, with amplitudes of $B_{g_1}$ and $B_{g_2}$, respectively. These functions are included in the formulations of $g_1(t)$ and $g_2(t)$ to simulate deviations from perfect sinusoidal oscillations of $g_1(t)$ and $g_2(t)$, aiming to reflect the characteristics of the experimentally measured data. In Eqs.~(\ref{eq:g1})~and~(\ref{eq:g2}), a relatively large range of variation (0--2) was considered for $B_{g_1}$ and $B_{g_2}$ to study the performance of the developed framework. The power spectrum densities of $g_1$ (solid curves) and $g_2$ (dotted curves) for $B_{g_1}=B_{g_2}=0,~0.5,~1,~1.5$,~and~2 were calculated and presented in Fig.~\ref{fig:PSDg1g2}. The time period utilized to obtain these power spectrum densities was considered to be 0.5~s similar to the duration of the data collection for the separate chemiluminescence measurements. The results in Fig.~\ref{fig:PSDg1g2} are stepped by $10^5$ for clarity purposes. As can be seen, for all tested background amplitudes, the PSDs of $g_1$ and $g_2$ feature relatively large amplitudes at the matching frequency.

\begin{figure*}[!t]
	\centering
	\includegraphics[width = 0.8\textwidth]{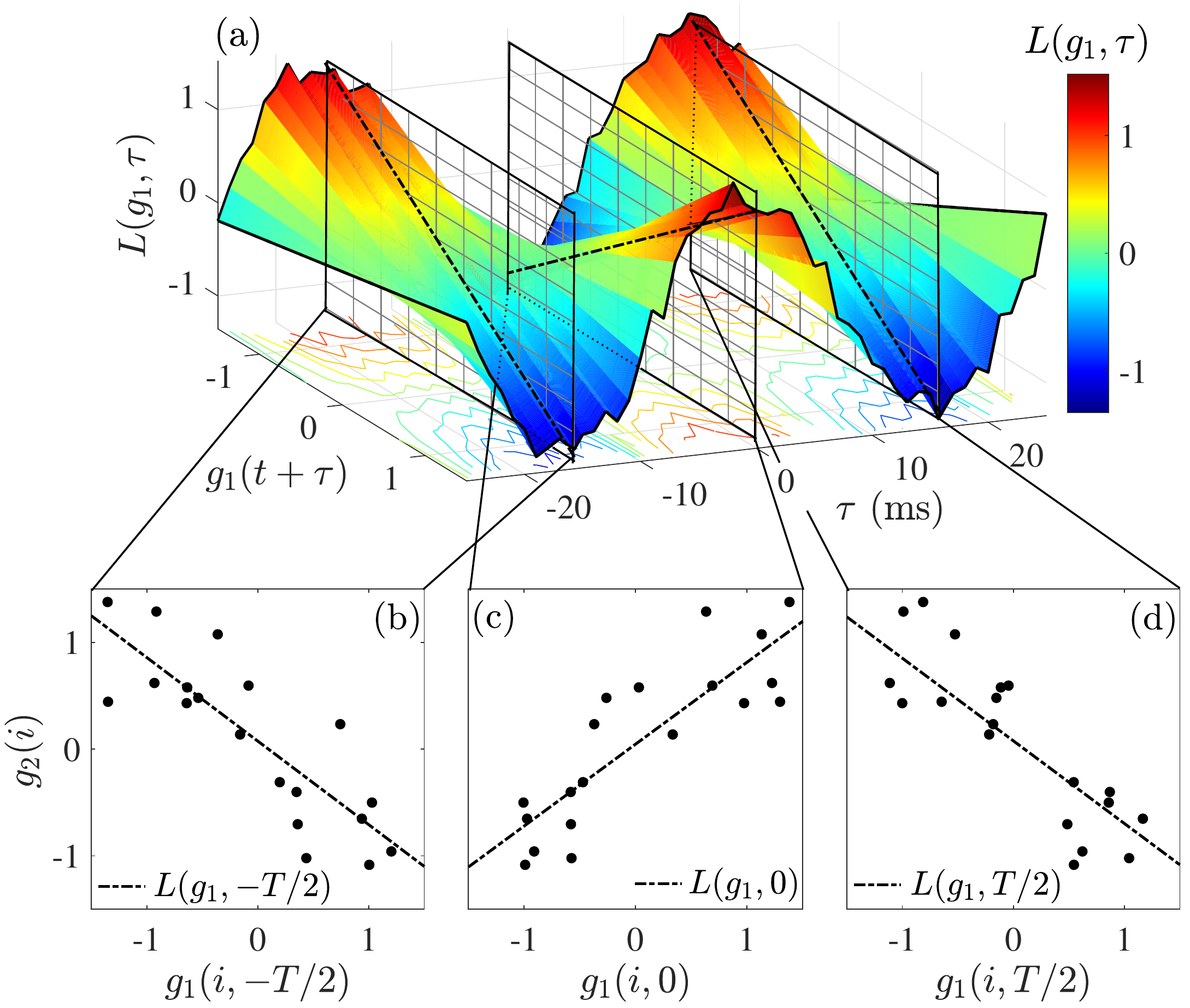}
	\caption{(a) The 3D variation of $L(g_1,\tau)$. (b--d) are $g_2(i)$ versus $g_1(i,-T/2)$, $g_1(i,0)$, and $g_1(i,T/2)$, respectively. The circular data points in (b--d) correspond to $p =19$ datasets and the dotted-dashed lines are the linear fits.}
	\label{fig:3Dfits}
\end{figure*}

For the analyses presented in this section, two number of training datasets ($p = 19$ and 199) were considered. $p = 19$ corresponds to the simultaneous experimental measurements performed in this study, with the acquisition timing shown in Fig.~\ref{fig:timing}. For $p=19$, 20 datasets were collected: 19 datasets are used for the training and one dataset is used for the prediction. One order of magnitude larger number of datasets is also considered to study the prediction accuracy of the framework: $p= 199$ for training and one dataset for prediction. The time duration for generating $g_1(t)$ and $g_2(t)$ corresponds to that shown in Fig.~\ref{fig:timing} and equals $p\Delta t_\mathrm{M}+\Delta t_\mathrm{CL}$, which is the summation of time periods $p\Delta t_\mathrm{M}$ (time period between the $p+1$ datasets), $\Delta t_\mathrm{CL}/2$ at the beginning of the data collection, and $\Delta t_\mathrm{CL}/2$ at the end of the data collection, with $\Delta t_\mathrm{M}=5$~s and $\Delta t_\mathrm{CL} = 50~\mathrm{ms}$ (see Fig.~\ref{fig:timing}). Thus, for prediction purposes, the time durations for generating $g_1(t)$ and $g_2(t)$ signals are 95.05 and 995.05 for $p = 19$ and 199, respectively. Following the simultaneous measurements performed in this study (see Fig.~\ref{fig:timing}), the values of $g_2(t)$ at $t_0+(i-1) \Delta t_\mathrm{M}$ are used for the prediction purposes, similar to the Mie scattering data, which are experimentally available at discrete times. $t_0$ is a reference time, which was set as half of the duration of the chemiluminescence datasets ($t_0 = \Delta t_\mathrm{CL}/2 = 25~\mathrm{ms}$). $i$ corresponds to the $i^\mathrm{th}$ dataset and changes from 1 to $p+1$. Since the chemiluminescence data is mimicked by $g_1(t)$, its variation is known at $t_0+(i-1) \Delta t_\mathrm{M}+\tau$, with $\tau$ being a time-lag and $-25~\mathrm{ms} \leq \tau \leq 25~\mathrm{ms}$, see Fig.~\ref{fig:timing}. Substituting $t_0+(i-1) \Delta t_\mathrm{M}+\tau$ and $t_0+(i-1) \Delta t_\mathrm{M}$ in Eqs.~(\ref{eq:g1})~and~(\ref{eq:g2}) for time, it is obtained that
 	
\begin{subequations}	
	\begin{equation}
		\label{eq:g1_bs}
		g_1(t_0+(i-1) \Delta t_\mathrm{M}+\tau) = \sin \left(2\pi f (t_0+(i-1) \Delta t_\mathrm{M}+\tau) \right)+B_{g_1} \mathcal{R}_1\left(t_0+(i-1) \Delta t_\mathrm{M}+\tau \right),
	\end{equation}
	\begin{equation}
		\label{eq:g2_bs}
		g_2(t_0+(i-1) \Delta t_\mathrm{M}) = \sin \left(2\pi f (t_0+(i-1) \Delta t_\mathrm{M}) \right)+B_{g_2} \mathcal{R}_2\left(t_0+(i-1) \Delta t_\mathrm{M} \right).
	\end{equation}
\end{subequations}	 
Since $\Delta t_\mathrm{M}$ and $t_0$ are fixed in the present study, $g_1(t_0+(i-1) \Delta t_\mathrm{M}+\tau)$ and $g_2(t_0+(i-1) \Delta t_\mathrm{M})$ are presented as $g_1(i,\tau)$ and $g_2(i)$, respectively. Thus, Eqs.~(\ref{eq:g1_bs})~and~(\ref{eq:g1_bs}) can be formulated as

\begin{subequations}	
	\begin{equation}
		\label{eq:g1_s}
		g_1(i,\tau) = \sin \left(2\pi f (t_0+(i-1) \Delta t_\mathrm{M}+\tau) \right)+B_{g_1} \mathcal{R}_1\left(t_0+(i-1) \Delta t_\mathrm{M}+\tau \right),
	\end{equation}
	\begin{equation}
		\label{eq:g2_s}
		g_2(i) = \sin \left(2\pi f (t_0+(i-1) \Delta t_\mathrm{M}) \right)+B_{g_2} \mathcal{R}_2\left(t_0+(i-1) \Delta t_\mathrm{M} \right).
	\end{equation}
\end{subequations}	 
The time-lag in Eq.~(\ref{eq:g1_s}) is a key parameter for the prediction framework developed here. First, for a given $\tau$, the values of $g_2(i)$ versus $g_1(i,\tau)$ (with $i \in [1,2,\cdots,p]$) were calculated. Then, for each value of $\tau$, least square fits to the above variations were obtained, with the corresponding linear fit shown in Fig.~\ref{fig:3Dfits}(a). This fit relates the values of $g_1$ measured at $t+\tau$ to $g_2$ measured at $t$. The 3D time-lag based formulation is given by 
\begin{equation}
	\label{eq:L}
	L(g_1(t+\tau),\tau) = a(\tau)+b(\tau)g_1(t+\tau).
\end{equation}
In Eq.~(\ref{eq:L}), $a(\tau)$ and $b(\tau)$ are the abscissa and 
slopes of the linear fit at a given time-lag. In Fig.~\ref{fig:3Dfits}(a), the 3D surface, $L(g_1,\tau)$, as well as the contours of $L(g_1,\tau)$ versus $\tau$ are colored based on the values of $L(g_1,\tau)$. Please note that, in the analyses, $\tau$ was varied in steps of $0.1$~ms to generate $L(g_1,\tau)$; however, only for clarity of presentation, $\tau$ was varied in steps of 1~ms to produce the 3D surface shown in Fig.~\ref{fig:3Dfits}(a). Also, for the results presented in Fig.~\ref{fig:3Dfits}, $B_{g_1} = B_{g_2} = 0.5$ and $p =19$. Variations of $g_2(i)$ versus $g_1(i,\tau$) for $\tau = -T/2$, 0, and $T/2$ are presented in Fig.~\ref{fig:3Dfits}(b--d), respectively. Here, $T$ is one period of oscillations and $T/2 = 1/(2f) = 15.6$~ms. The results in Fig.~\ref{fig:3Dfits}(c) suggest that the slope of the linear fit at $\tau = 0$ is positive; however, it becomes negative at $\tau = -15.6$ and 15.6~ms. In fact, the slope of the linear fit varies as a cosine-shaped function, which is due to the time-lag between $g_1(t+\tau)$ and $g_2(t)$ and the sinusoidal functions in Eqs.~(\ref{eq:g1_s})~and~(\ref{eq:g2_s}).

\begin{figure*}[!t]
	\centering
	\includegraphics[width = 1\textwidth]{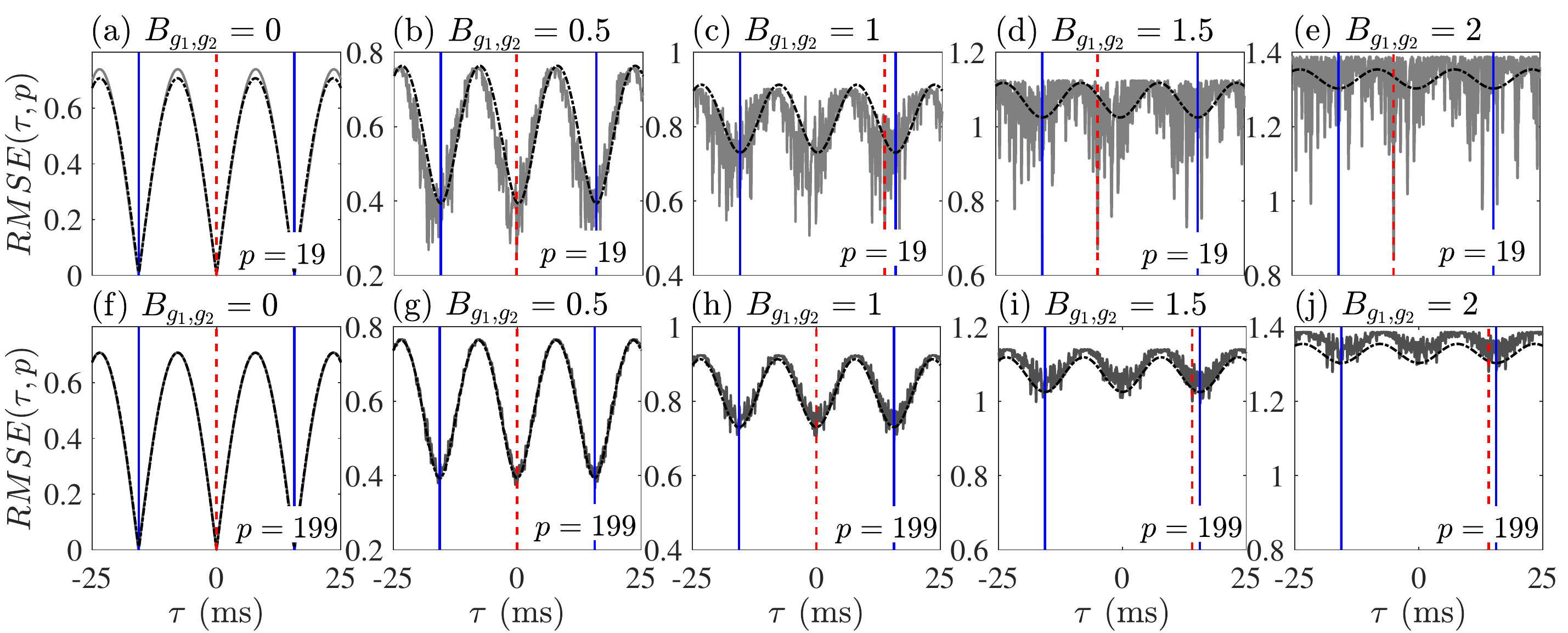}
	\caption{Variation of $RMSE(\tau,p)$ for 19 (first row) and 199 (second row) training datasets, respectively. The first to fifth columns pertain to $B_{g_1} = B_{g_2} = 0$, 0.5, 1, 1.5, and 2.0, respectively. The black dotted-dashed curves are the right-hand-side of Eq.~(S.8) from Appendix~C. The red dashed line corresponds to the time-lag at which $RMSE$ is minimized. The blue dashed lines highlight $\tau = \pm 1/(2f)$.}
	\label{fig:error}
\end{figure*}

After estimating $L(g_1,\tau)$ (which is a time-lag dependent and linear fit for estimation of $g_2(i)$), the root-mean-square-error ($RMSE$) was calculated to quantify the accuracy in predicting $g_2(i)$ using $g_1(i,\tau)$ at $t = t_0 + (i-1) \Delta t_\mathrm{M} + \tau$. $RMSE$ is given by
\begin{equation}
	\label{eq:RMSE}
	RMSE(\tau,p) = \sqrt{\frac{\sum_{i = 1}^{i = p} \left[g_2(i)-L(g_1,\tau)\right]^2}{p}}.
\end{equation}
Figure~\ref{fig:error} presents the variation of $RMSE$ versus $\tau$ for $p = 19$ (first row) and $p = 199$ (second row). The first to fifth columns pertain to $B_{g_1} = B_{g_2} = 0$, 0.5, 1, 1.5, and 2.0, respectively. The time-lag within $-1/(2f) < \tau < 1/(2f)$ that $RMSE$ minimizes is referred to as $\tau^*$. This is the time-lag at which $L(g_1,\tau^*)$ provides the most accurate prediction of $g_2$. Thus, the following equation can be used for approximating $g_2$.
\begin{equation}
	\label{eq:g2Pred_p}
	g_2(t) \approx L(g_1,\tau^*) = a(\tau^*)+b(\tau^*)g_1(t+\tau^*),
\end{equation}
where $a(\tau^*)$ and $b(\tau^*)$ are the abscissa and slopes of the linear fit at $\tau=\tau^*$. In Fig.~\ref{fig:error}, $\tau = \pm 1/(2f)$ are shown by the solid blue lines for clarity. $\tau^*$ for all examined background values and for both $p=19$ and $p=199$ are obtained and shown by the red dashed lines in Fig.~\ref{fig:error}. For $g_1$ and $g_2$ used in Eqs.~(\ref{eq:g1})~and~(\ref{eq:g2}), the true value of $\tau^*$ is zero. 

The results in Fig.~\ref{fig:error} show that, for $p=19$, the correct value of $\tau^*$ is predicted for $B_{g_1}=B_{g_2}=0$ and 0.5. However, for $B_{g_1}=B_{g_2} \geq 1$, the time-lag that leads to the minimum value of $RMSE$ is estimated to be non-zero for $p=19$. It can be seen that increasing the number of training datasets from $p = 19$ to $p=199$ increases the background value for which $\tau^*$ is correctly estimated from 0.5 to 1. However, for $p = 199$, the correct value of $\tau^*$ could not be obtained for $B_{g_1}=B_{g_2} = 1.5$ and 2.0. For $p =\infty$, Eqs.~(\ref{eq:g1_s},~\ref{eq:g2_s},~and~\ref{eq:RMSE}) were utilized and an analytical formulation for $RMSE$ versus $\tau$ was obtained, see Appendix~C. The variations of $RMSE(\tau,\infty)$ are shown by the dotted-dashed black curves in Fig.~\ref{fig:error}. As can be seen, independent of the background value, the correct value of the time-lag can be estimated as the number of training datasets approaches infinity. It is also obtained that the estimated $RMSE$ for $p=19$ (which is relevant to the experiments) is close to that for $p=\infty$ provided $B_{g_1} = B_{g_2} \leq 0.5$. Though not presented here, the performance of the framework for periodic functions with non-zero time-lag was examined, and it was obtained that the developed framework can allow for estimating the true value of $\tau^*$. 

\begin{figure*}[!t]
	\centering
	\includegraphics[width = 1\textwidth]{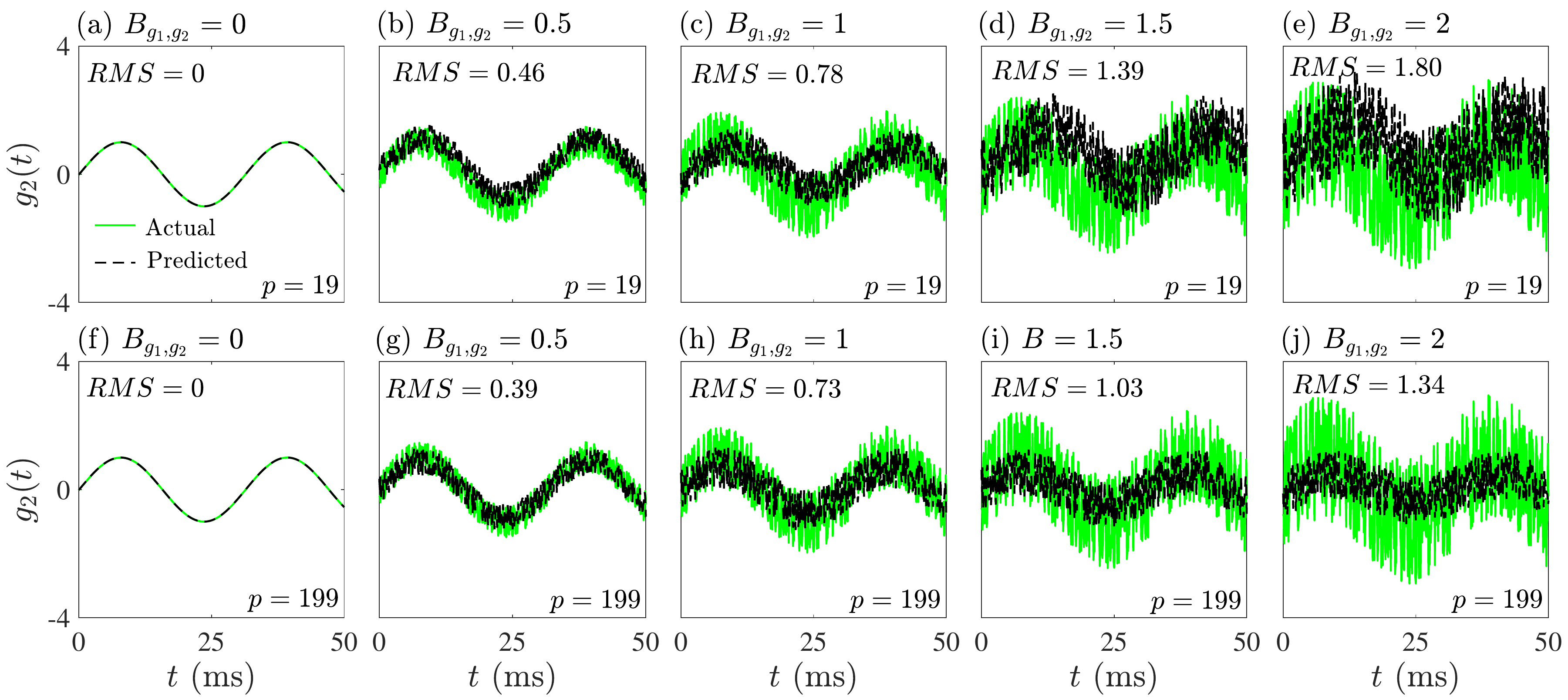}
	\caption{The actual (solid green) versus predicted (dashed black) values of $g_2(t)$. The first and second rows pertain to 19 and 199 training datasets, respectively. The first to fifth columns are for amplitudes of the random background signals equal to 0, 0.5, 1, 1.5, and 2.0, respectively.}
	\label{fig:time-resolved predictions}
\end{figure*}

Using $\tau^*$, $g_1(t)$, and Eq.~(\ref{eq:g2Pred_p}), the predicted temporal variation of $g_2(t)$, which is $L(g_1,\tau^*)$ was obtained. Figure~\ref{fig:time-resolved predictions} presents the temporal variations of actual $g_2(t)$ and predicted $g_2(t)$ in solid green and dashed black lines, respectively. The first and second rows pertain to relatively small ($p = 19$) and large ($p = 199$) numbers of training datasets. The results in the first to fifth columns pertain to increasing values of the background signal amplitude. $g_1(t)$ and $g_2(t)$ were generated for a relatively long period of time, and for presentation purposes, the results in Fig.~\ref{fig:time-resolved predictions} are only shown for the time period of 50~ms. Similar results are obtained for other time durations. Overlaid on the figure are the RMS of the predicted signal subtracted from the actual signal. As can be seen in Fig.~\mbox{\ref{fig:time-resolved predictions}}, for a given value of the background, increasing $p$ decreases the RMS of the differences between actual $g_2(t)$ and the predicted $g_2(t)$. It can also be seen that, for $B_{g_1}=B_{g_2}=0$ and 0.5, the prediction of the framework is close to the actual signal. It is important to highlight that, for the large background amplitude values ($B_{g_1}=B_{g_2} \geq 1$) and for the smaller value of the training dataset, the predicted signal is time lagged, leading to incorrect prediction of $g_2(t)$, compare the black and green variations in Fig.~\ref{fig:time-resolved predictions}(c--e). This is due to incorrect estimation of $\tau^*$, as shown in Fig.~\ref{fig:error}(c--e) for large background values.

\begin{figure*}[!t]
	\centering
	\includegraphics[width = 1\textwidth]{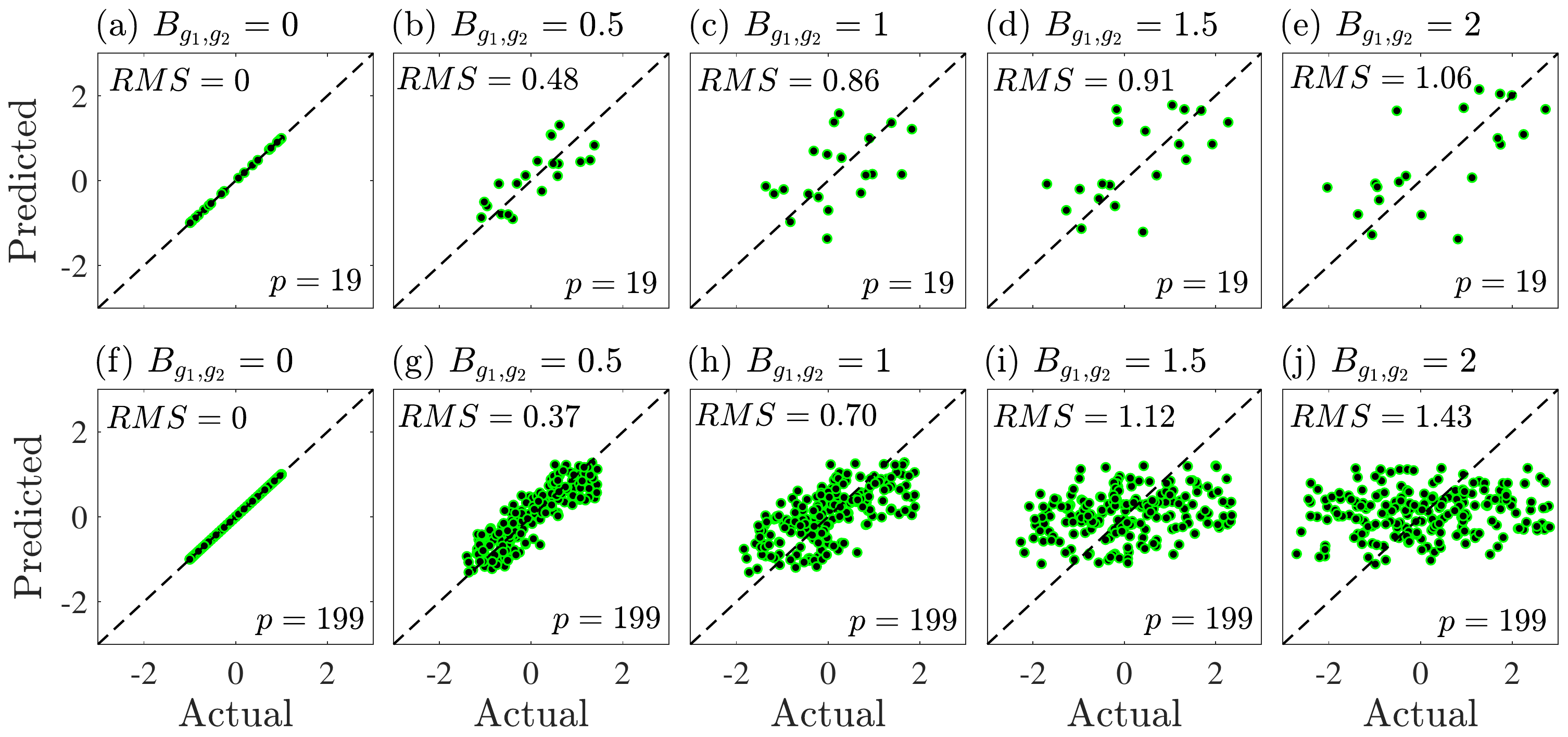}
	\caption{Predicted versus actual values of $g_2(i)$ for $p = 19$ (first row) and $p=199$ (second row). The first to fifth columns pertain to $B_{g_1} = B_{g_2} = 0$, 0.5, 1.0, 1.5, and 2.0, respectively.}
	\label{fig:predictionsmodel}
\end{figure*}

The results discussed above suggest that the presence of a large amplitude background signal can lead to an incorrect prediction of $\tau^*$ and as a result an incorrect prediction of $g_2$. In order to study the effect of inaccurately estimating $\tau^*$ on the prediction of $g_2$, all possible combinations of $p$ training datasets from $p+1$ available datasets were considered to predict $g_2(i)$ using Eq.~(\ref{eq:g2Pred_p}). That is, in addition to using datasets $[1, 2, \cdots, p]$ for predicting $g_2(i)$ for the $(p+1)^\mathrm{th}$ dataset, the rest of the combinations, for example datasets of $[1, 2, \cdots, p-1, p+1]$, were considered to predict $g_2(i)$ for the $p^\mathrm{th}$ excluded dataset. Predicted versus the actual values of $g_2(i)$ for $i=1$ to $p+1$ are presented in Fig.~\ref{fig:predictionsmodel}. Also, presented in the figure are the RMS of the predicted results subtracted by the actual results. As can be seen, increasing $p = 19$ to 199 decreases the $RMS$ for $B_{g_1}=B_{g_2} \leq 1$. For $B_{g_1}=B_{g_2} \geq 1.5$, however, the estimated $RMS$ for both $p = 19$ and $p = 199$ are relatively large. Overall, the linear-regression training model utilized in the present study suggests that, for signals with characteristics similar to those discussed in this section, $p=19$ datasets are sufficient for the prediction of the twentieth dataset for the random variation background amplitude of about 50\%, which is a relatively large background amplitude. The developed model here is used in the next section for prediction purposes.

\section{Results}
\label{Results}

\begin{figure*}[!t]
	\centering
	\includegraphics[width = 1\textwidth]{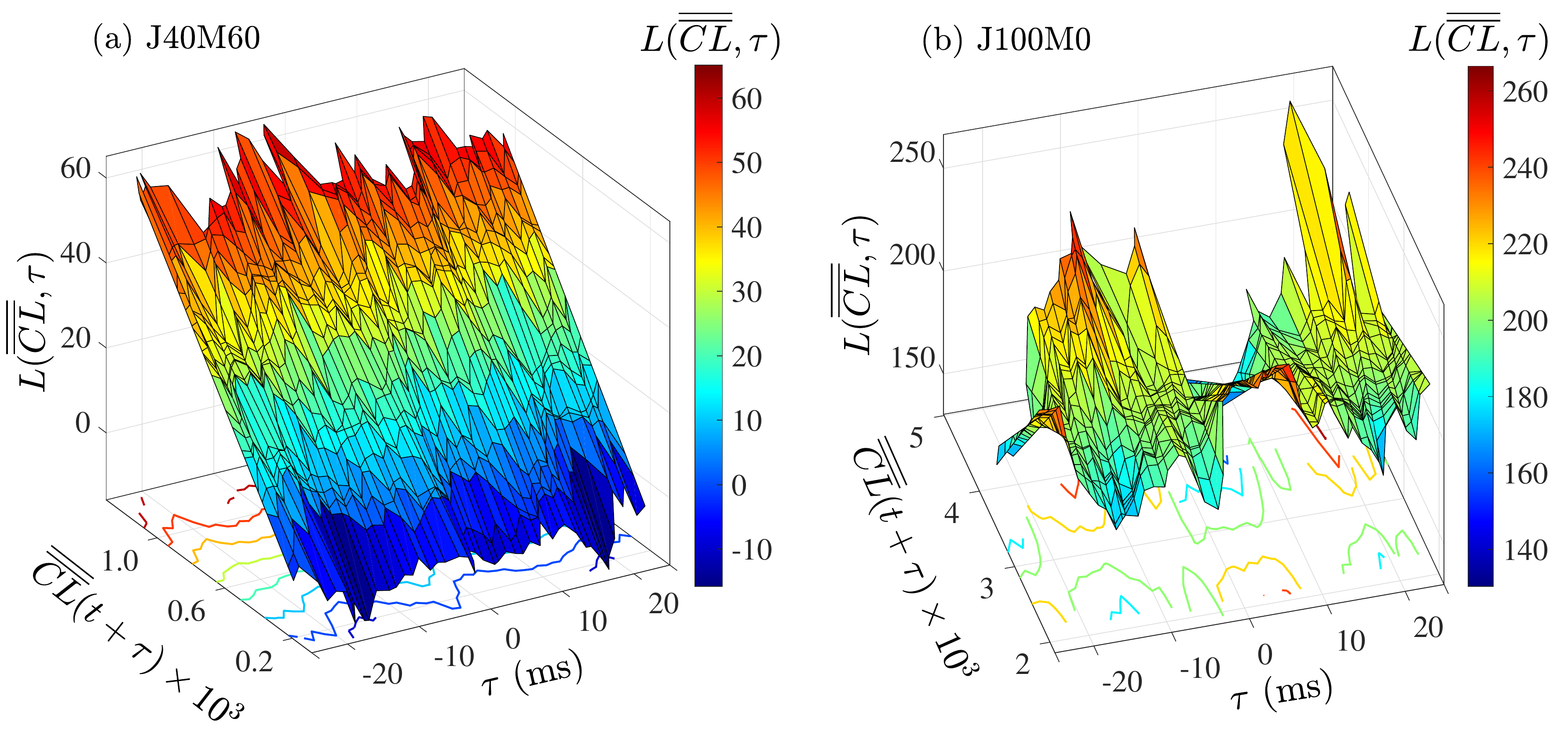}
	\caption{$L(\overline{\overline{CL}},\tau)$ for (a) J40M60 and (b) J100M0, respectively.}
	\label{fig:Realdata3D}
\end{figure*}

The framework developed in section~\ref{framework} is employed for the time-resolved prediction of the spray number of droplets within the plane of illumination using the time-resolved and spatially averaged flame chemiluminescence ($\overline{\overline{CL}}$) and sparse Mie scattering data. Following section~\ref{framework}, $g_1(t) = \overline{\overline{CL}}(t)$ and $g_2(t) = n(t)$. First, the 3D variations of $L(g_1,\tau)$ were obtained and presented in Figs.~\ref{fig:Realdata3D}(a) and (b) for the test conditions J40M60 and J100M0, respectively. Please note that 19 training datasets are used to perform this analysis and the 20$^\mathrm{th}$ dataset is utilized for the prediction. For J100M0, the 3D surface presented in Fig.~\ref{fig:Realdata3D}(b) and the contours of $L(g_1,\tau)$ feature a nearly periodic variation, which is similar to that shown in Fig.~\ref{fig:3Dfits}. However, for J40M60, the results in Fig.~\ref{fig:Realdata3D}(a) do not feature the periodic variations. This is because the matching frequency band of the spatially averaged flame chemiluminescence and spray is smaller than 10~Hz, which corresponds to a time period larger than 100~ms, and as a result, one complete cycle of oscillation cannot be observed in the 3D presentation. Nonetheless, as will be shown and discussed later, the above framework can allow for predicting the number of droplets for both test conditions J40M60 and J100M0.

\begin{figure*}[!t]
	\centering
	\includegraphics[width = 0.95\textwidth]{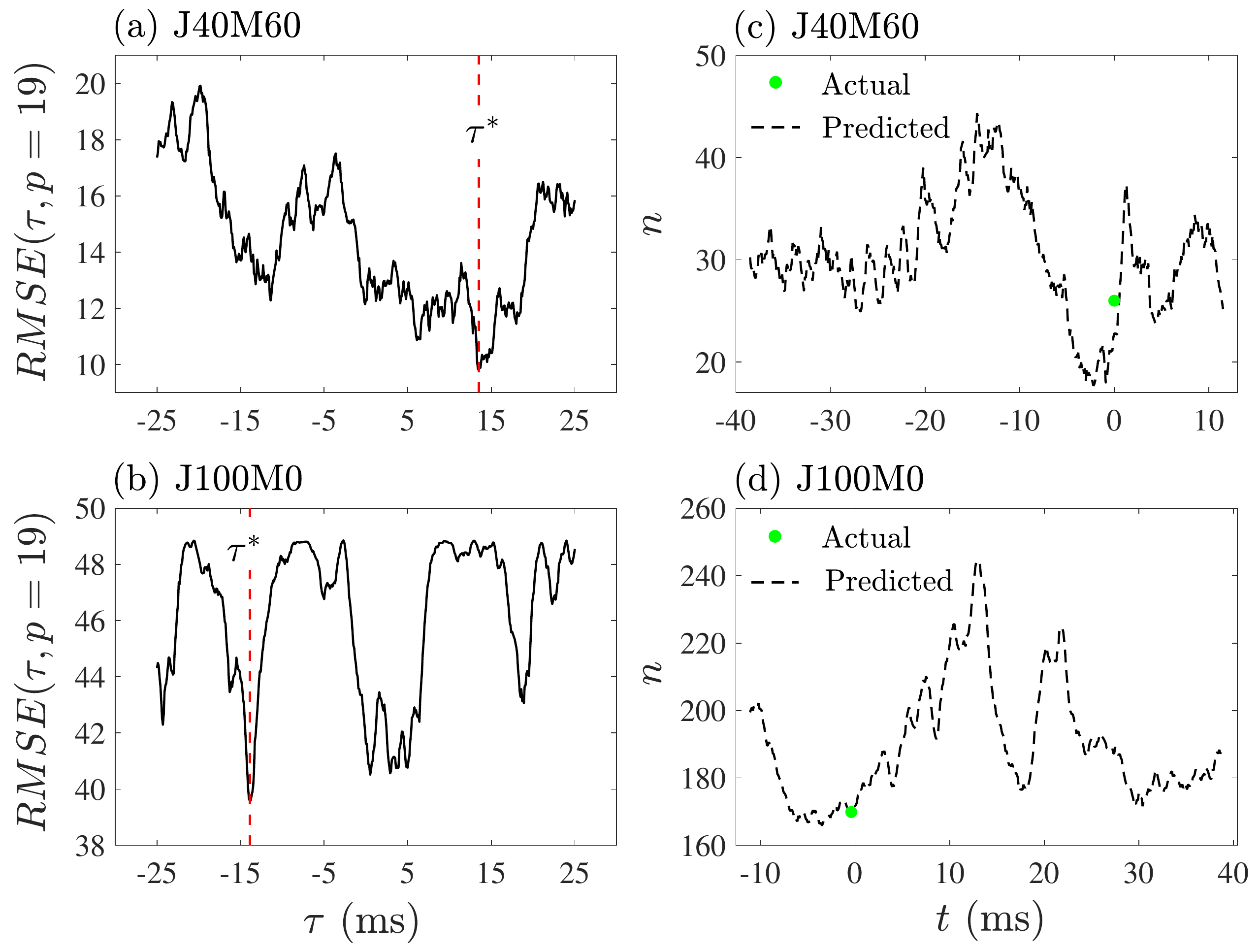}
	\caption{(a) and (b) are variations of $RMSE(\tau,p=19)$ for J40M60 and J100M0, respectively. The time-lag at which $RMSE$ minimizes is shown in (a and b) by the red dashed lines. (c) and (d) are the predicted (black dashed lines) variation and measured (green circular data symbol) value of the spray number of droplets for the 20$^\mathrm{th}$ dataset of test conditions J40M60 and J100M0, respectively.}
	\label{fig:RealdataRMSE}
\end{figure*}

Figure~\ref{fig:RealdataRMSE}(a) and (b) present the variation of $RMSE$ obtained from Eq.~(\ref{eq:RMSE}) versus $\tau$ for J40M60 and J100M0, respectively. $\tau^*$ for both conditions are identified by the red dashed lines. Specifically, $\tau^*$ equals 13.5~ms and -13.9~ms for J40M60 and J100M0, respectively. Following the framework discussed in section~\ref{framework}, the estimated values of $\tau^*$ in Figs.~\ref{fig:RealdataRMSE}(a) and (b) were used to obtain the time-resolved variation of $n(t)$ for the 20$^\mathrm{th}$ dataset from Eq.~(\ref{eq:g2Pred_p}). These variations are presented in Figs.~\ref{fig:RealdataRMSE}(c) and (d) for J40M60 and J100M0, respectively. In these figures, $t=0$~ms is set as the time at which the Mie scattering data ($n$) was collected for the 20$^\mathrm{th}$ dataset. From Eq.~(\ref{eq:g2Pred_p}), the argument of $g_1$, which is $t+\tau^*$, varies between -25~ms and 25~ms. Thus, the argument of $g_2$, which is $t$, varies between $-25-\tau^*$~ms and $25-\tau^*$~ms. Also overlaid by the circular green data point on Figs.~\ref{fig:RealdataRMSE}(c) and (d) are the actual values of $n$. As can be seen, the developed framework in section~\ref{framework} can accurately predict the actual value of $n$ at $t=0$~ms and for both test conditions. 

\begin{figure*}[!t]
	\centering
\includegraphics[width = 0.5\textwidth]{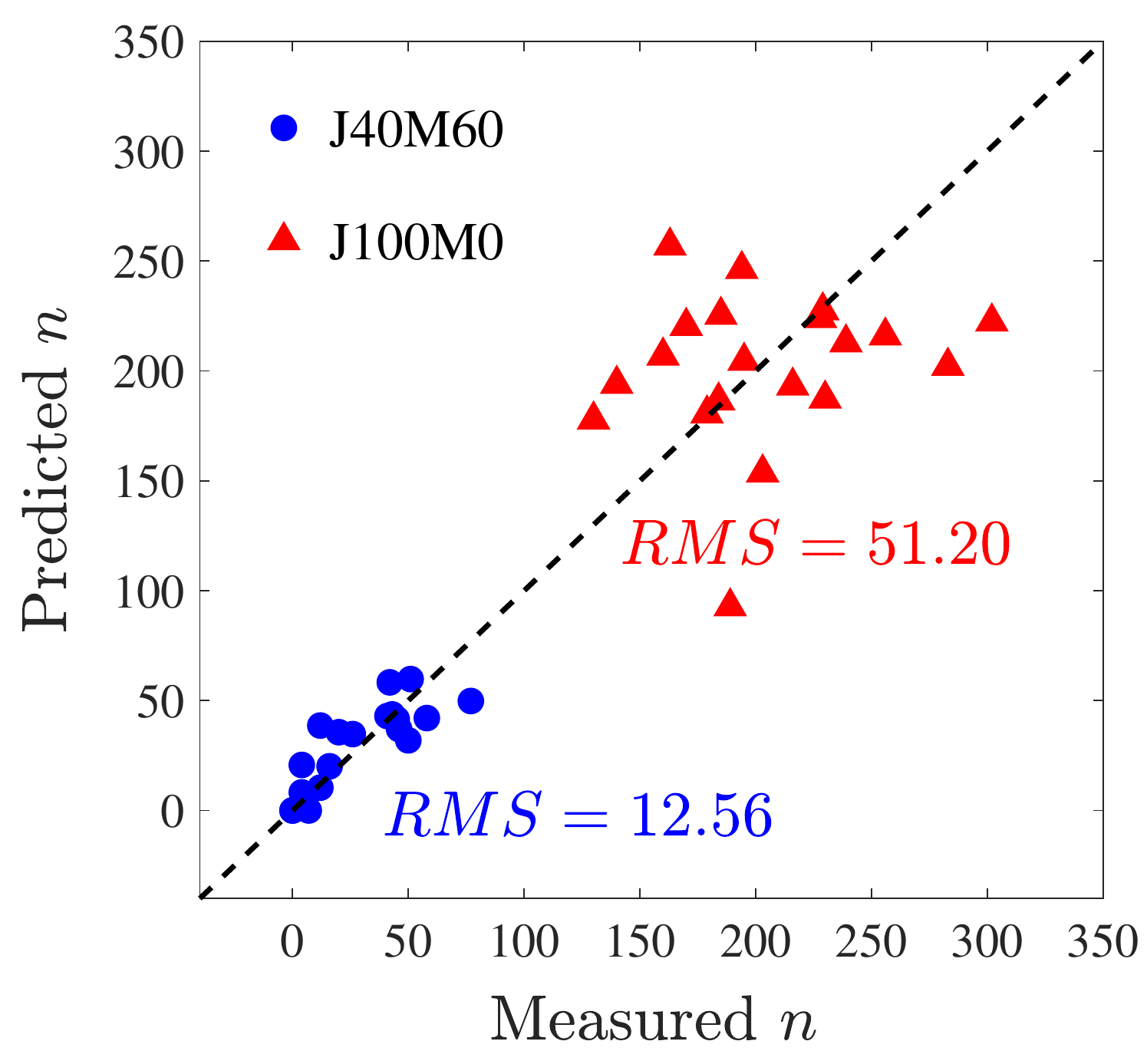}
	\caption{Predicted versus measured number of droplets, $n(i)$.}
	\label{fig:Realdatascatterplots}
\end{figure*}

In order to assess the accuracy of the predictions for the number of droplets, all 20 collected datasets were used and the predicted values of $n$ versus the corresponding measured values are presented by the blue circular (for J40M60) and red triangular (for J100M0) data symbols in Fig.~\ref{fig:Realdatascatterplots}, respectively. The procedure followed to obtain the results in Fig.~\ref{fig:Realdatascatterplots} is identical to that used for obtaining those in Fig.~\ref{fig:predictionsmodel}. The corresponding values of the RMS of the difference between the measured and predicted data are also presented in the figure. The RMS values are about 13 and 51 droplets for J40M60 and J100M0, respectively. 

In section~\ref{framework}, the number of droplets and the mass of the droplets were estimated using the shadowgraphy technique, and it was shown that the PSD of these parameters feature large amplitude oscillations at frequencies close to those of the flame chemiluminescence. Thus, it was assumed that the number of droplets measured inside a plane using the Mie scattering data features matching frequencies to that of the flame chemiluminescence. Comparison of the actual fuel number of droplets and the predicted values, see Fig.~\ref{fig:predictionsmodel}, suggest the above assumption along with the developed framework allow for the prediction of the fuel number of droplets. In essence, the discussions presented here show that using the knowledge related to the coupling between the spray and the flame chemiluminescence along with the framework developed here, the time-resolved variation of spray can be obtained from the corresponding sparsely measured data. In our future work, the above developed and tested framework will be used to provide improved understanding related to the dynamics of 2D measured flame structure and its interaction with the fuel droplets.

\section{Conclusions}
\label{Conclusions}
The temporal variation of Jet A-1 spray number of droplets inside a plane was predicted using the time-resolved flame chemiluminescence, sparse Mie scattering, as well as a physics-informed data-driven framework, which was developed in this study. The spray flames were characterized using separate flame chemiluminescence, interferometric laser imaging for droplet sizing, shadowgraphy, plenum pressure measurements. Also, simultaneous flame chemiluminescence, Mie scattering, and pressure data were collected for 20 datasets for the purposes of the above framework development and for prediction. For the simultaneous measurements, the acquisition frequency of the pressure, flame chemiluminescence, and Mie scattering were set to 100000, 10000, and 0.2 Hz, respectively. A gas turbine model combustor was retrofitted to operate with both Jet A-1 and a mixture of methane and air. The combustor was operated at a fixed power of 10~kW. The flow rates of methane, Jet A-1, and air were adjusted to generate test conditions corresponding to: (i) methane and air perfectly premixed flames, (ii) Jet A-1 spray flames, and (iii) dual fuel flames with 40\% and 60\% of the power generated by Jet A-1 and methane, respectively.

For all test conditions, the plenum pressure features broadband oscillations near 300 to 700 Hz, which do not match those of the flame chemiluminescence. Specifically, for perfectly premixed flames, the spatially averaged flame chemiluminescence features a dominant frequency at about 100~Hz, which reduces to $\lesssim$~10~Hz and 10--40~Hz for the dual fuel and Jet A-1 flames, respectively. Analysis of both non-reacting and reacting shadowgraphy images confirmed that the number and mass of the droplets oscillations feature dominant frequencies that match those of the flame chemiluminescence both for dual fuel and Jet A-1 spray flames. This suggested that the flame chemiluminescence is driven by fuel injection instability.

A time-lag and linear regression-based analysis was used to develop a framework that allows to predict time-resolved variation of an objective signal using a time-resolved input signal and sparse information from the objective signal. In the present study, the objective and input signals are the number of spray droplets measured inside a plane from the Mie scattering and the spatially averaged flame chemiluminescence, respectively. The framework was developed assuming the objective and input signals feature relatively large amplitude oscillations at a matching frequency. In the developed framework, first, linear fits were used to obtain relations between the flame chemiluminescence and spray number of droplets for several time-lags. Then, the time-lag at which the error minimizes was obtained. Finally, the time-lag and the slope of the fits were used to predict the temporal variation of the spray number of droplets. Of importance for the development of the framework is the number of training datasets, whose influence on the accuracy of predictions was assessed. The developed framework was then used to predict the time-resolved variation of the number of spray droplets inside a plane. It was shown that the number of droplets predicted by the developed framework matches relatively well the measured number of droplets.

The framework developed and assessed in the present study allows for utilizing temporally resolved chemiluminescence data and sparse number of droplets to predict the temporal variation of the spray number of droplets. Such information is of importance for spray characterization and understanding the coupling between the spray and the flame. For experiments that the high-speed spray data is not available, the developed framework in this study can be used for predicting the missing information.

\section*{Acknowledgments}
The authors are grateful for the financial support from the Natural Sciences Engineering Research Council of Canada and Zentek through the Alliance grant ALLRP 567111-21 as well as MITACS and Machinery Analytics through grant IT25776. The authors acknowledge KF Aerospace for providing Jet A-1.

\section*{Appendix A: Jet A-1 volume flow rate calibration}
\label{AppendixA}
\begin{figure*}[!t]
	\centering
	\includegraphics[width = 0.5\textwidth]{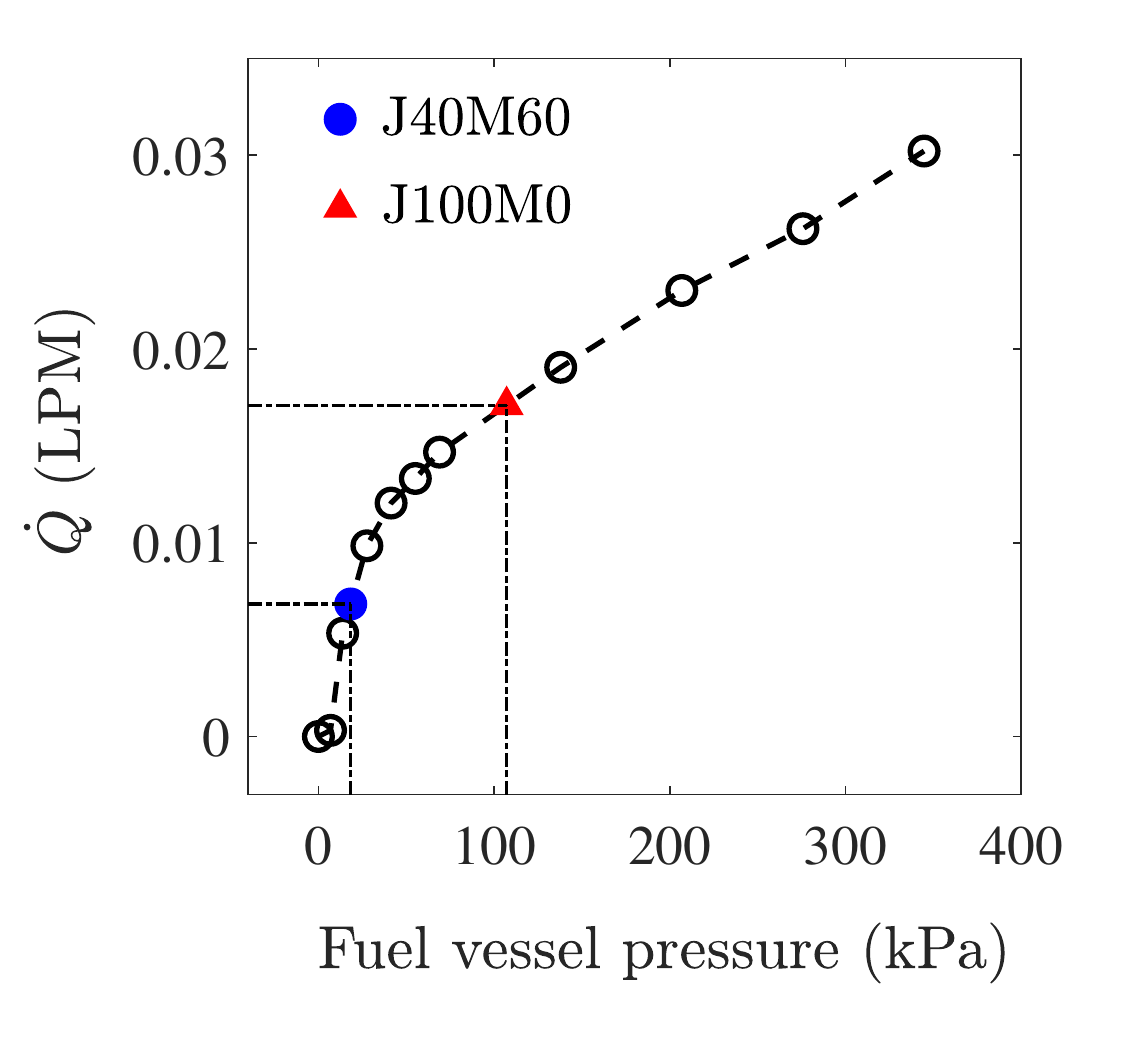}
	\caption{The relation between the volumetric flow rate of Jet A-1 and the liquid fuel vessel pressure.}
	\label{fig:Calibration}
\end{figure*}

For the Jet A-1 flow rate calibration, the combustion chamber shown in Fig.~\ref{fig:burner} was removed, the burner was flipped vertically, and it was connected to a sealed container, whose weight was monitored. Care was taken not to alter the fuel delivery system during this procedure. Then, the dual valve pressure-controller (item~3 in Fig.~\ref{fig:diagnostics}) was used to set the nitrogen pressure to several values ranging from about 0 to 345~kPa. For each pressure, the injector operated for 300~s and the weight of the collected Jet A-1 was measured. Using the measured density of Jet A-1 (812.0~$\mathrm{kg/m^3}$), the volume of the collected liquid was obtained (in liters) and divided by the duration (300~s) to calculate the fuel volume flow rate ($\dot{Q}$). Figure~\ref{fig:Calibration} presents the variation of $\dot{Q}$ versus the vessel pressure. J40M60 and J100M0 test conditions are also overlaid by the solid blue circle and red triangular symbols, respectively. It was obtained that the vessel gauge pressures of 18.4 and 107.1 kPa lead to $\dot{Q} = 6.8$ and 17.1 cubic centimeters per minute (5.6 and 13.9 $\mathrm{gr/min}$ as tabulated in Table~\ref{Tab:Testedconditions}), respectively.

\section*{Appendix B: Resolutions of the utilized optical diagnostics}
\label{AppendixB}

\begin{figure*}[!t]
	\centering
	\includegraphics[width = 0.6\textwidth]{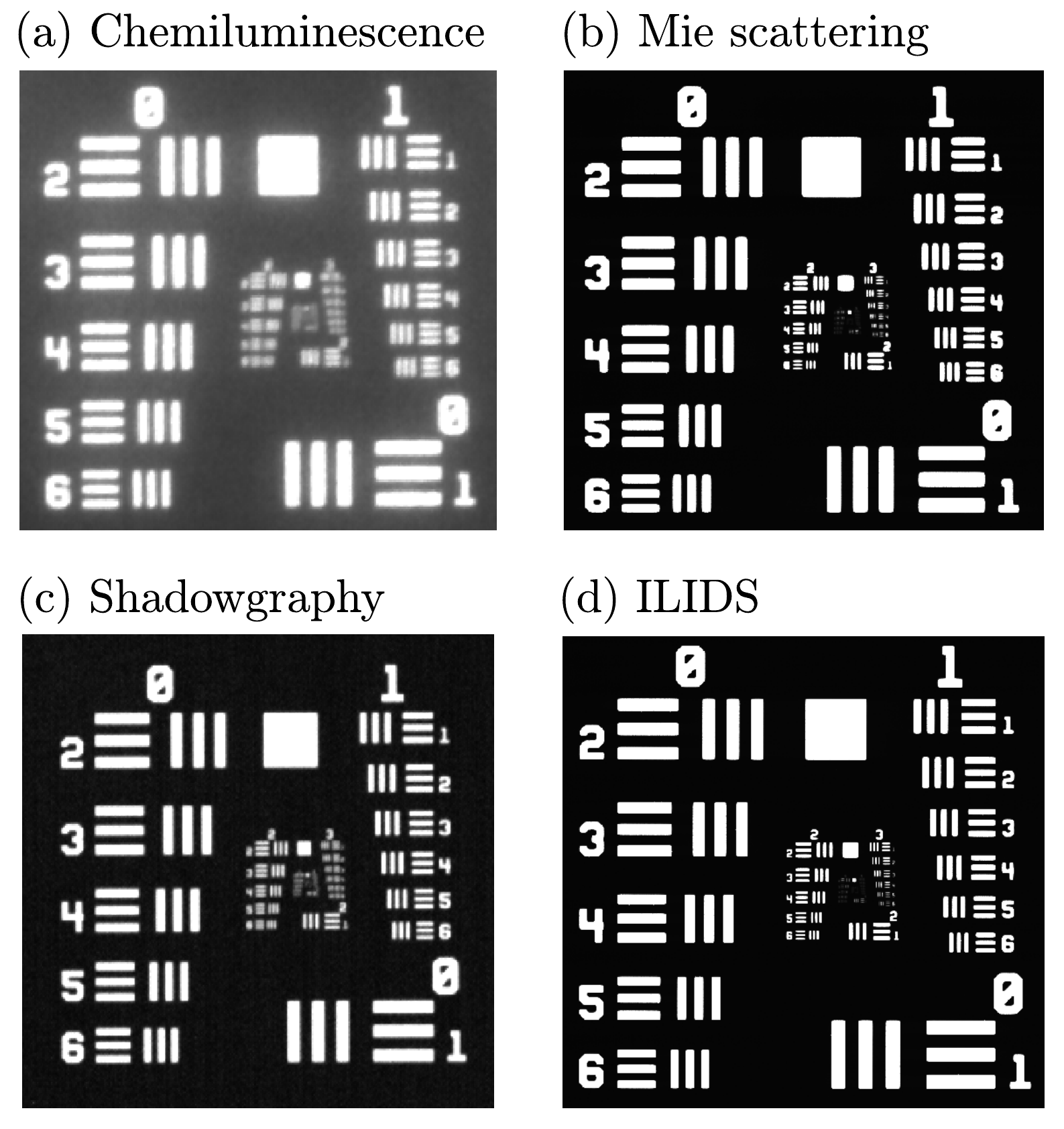}
	\caption{USAF 1951 images acquired using (a) chemiluminescence, (b) Mie scattering, (c) shadowgraphy, and (d) ILIDS diagnostics.}
	\label{fig:ResVis}
\end{figure*}

\begin{figure*}[!t]
	\centering
	\includegraphics[width = 1\textwidth]{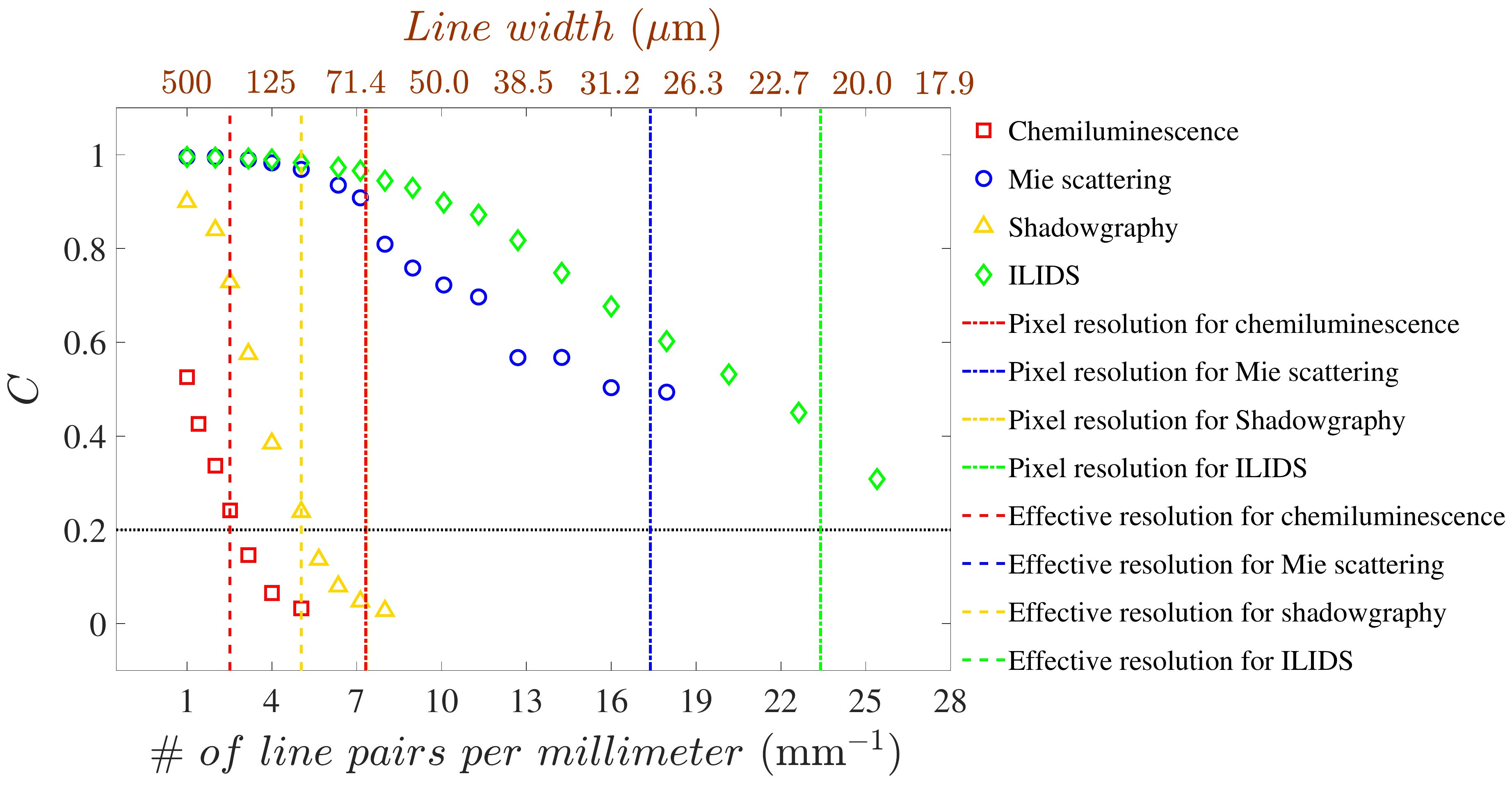}
	\caption{Contrast and resolution for the utilized optical diagnostics.}
	\label{fig:Res}
\end{figure*}

Similar to~\cite{mosadegh2022graphene,papageorge2014recent}, the USAF~1951 target plate was used to determine the spatial resolution of the optical diagnostics used in this study. Figure~\ref{fig:ResVis} shows the image of the USAF plate captured by the flame chemiluminescence, Mie scattering, shadowgraphy, and ILIDS diagnostics. The pixel resolution for the flame chemiluminescence, Mie scattering, shadowgraphy, and ILIDS measurements were 68.4, 28.8, 68.0, and 21.4~$\mathrm{\mu}$m, respectively. These values were converted to the number of line pairs per millimeter and are shown by the red, blue, yellow, and green dotted-dashed lines in Fig.~\ref{fig:Res}. Following Refs.~\cite{mosadegh2022graphene,papageorge2014recent}, the image contrast was defined as
\begin{equation}
	\label{eq:Contrast}
	C=\frac{I_\mathrm{max}-I_\mathrm{min}}{I_\mathrm{max}+I_\mathrm{min}},
\end{equation}	
where $I_\mathrm{max}$ and $I_\mathrm{min}$ are the maximum and minimum light intensities acquired across each group of the USAF~1951 lines. $C$ was calculated for each group of the USAF~1951 lines using the chemiluminescence, Mie scattering, shadowgraphy, and ILIDS imaging systems, with the corresponding results shown in Fig.~\ref{fig:Res} by the red square, blue circle, yellow triangular, and green diamond shape data symbols, respectively. Similar to Refs.~\cite{mosadegh2022graphene,papageorge2014recent}, $C=0.2$ was used for determining the effective spatial resolution of the diagnostics. These effective resolutions are shown by the dashed lines in the figure. As shown in Fig.~\ref{fig:Res}, the effective spatial resolutions are 198.4~$\mathrm{\mu}$m and 99.2~$\mathrm{\mu}$m for the flame chemiluminescence and shadowgraphy imaging systems, respectively. However, for the Mie scattering and ILIDS, $C$ is larger than 0.2; and, the effective resolutions of these imaging techniques are taken to be identical to the corresponding pixel resolutions (28.8~$\mathrm{\mu}$m for the Mie scattering and 21.4~$\mathrm{\mu}$m for the ILIDS). 

For the ILIDS measurements, the above argument can be used to identify the range of the detectable droplets diameter. In our ILIDS measurements, the identified fringe patterns feature a diameter of about 2~mm, with a representative pattern shown in Fig.~\ref{fig:datareductionILIDS}. Thus, the maximum number of fringes that can be resolved in the ILIDS experiments is $2~\mathrm{mm}/21.4~\mu \mathrm{m} \approx 45$. Using Eq.~(\ref{eq:ILIDS}) along with the minimum number of resolved fringes (which is 2) as well as the maximum number of fringes, ILIDS allows for resolving droplets with diameter $4.8 \leq d \leq 107.2 ~\mathrm{\mu}$m. This means that the droplets with diameter larger than 107.2 cannot be detected using the ILIDS technique.

\section*{Appendix C: $RMSE(\tau,\infty)$ closed formulation}
\label{AppendixC}

For discrete time steps, Eq.~(\mbox{\ref{eq:L}}) can be written as
\begin{equation}
	\label{eq:L_discrete}
	L(g_1(i,\tau),\tau) = a(\tau)+b(\tau)g_1(i,\tau).
\end{equation}
Using Eq.~(\mbox{\ref{eq:L_discrete}}) and the Right-Hand-Side (RHS) of Eq.~(\ref{eq:RMSE}), $RMSE^2$ can be calculated from
\begin{equation}
	\label{eq:RMSE1}
	RMSE^2(\tau,p)=\frac{1}{p}\sum_{i=1}^{i=p}[g_2(i)-a(\tau)-b(\tau)g_1(i,\tau)]^2.
\end{equation}   
Then, substituting $g_1(i,\tau)$ and $g_2(i)$ from Eqs.~(\ref{eq:g1_s} and \ref{eq:g2_s}) in Eq.~(\ref{eq:RMSE1}), it is obtained that
\begin{multline}
	\label{eq:RMSE2}
	RMSE^2(\tau,p) = \frac{1}{p} \sum_{i=1}^{i=p} [\sin(2\pi f(t_0+(i-1)\Delta t_\mathrm{M})) +B_{g_2}\mathcal{R}_2(t_0+(i-1)\Delta t_\mathrm{M}) -a(\tau)\\ -b(\tau)\sin(2\pi f(t_0+(i-1)\Delta t_\mathrm{M}+\tau))-b(\tau)B_{g_1}\mathcal{R}_1(t_0+(i-1)\Delta t_\mathrm{M}+\tau)]^2.
\end{multline}   
Expanding the second order polynomial in the above equation, it can be shown that
\begin{multline}
	\label{eq:RMSE3}
	RMSE^2(\tau,p) = \underbrace{\frac{\sum_{i=1}^{i = p} \sin^2(2\pi f(t_0+(i-1)\Delta t_\mathrm{M}))}{p}}_{Term~1} + \underbrace{\frac{\sum_{i=1}^{i = p} B_{g_2}^2\mathcal{R}_2^2(t_0+(i-1)\Delta t_\mathrm{M})}{p}}_{Term~2} + \\ \underbrace{\frac{\sum_{i=1}^{i = p} a^2(\tau)}{p}}_{Term~3} + \underbrace{\frac{\sum_{i=1}^{i = p} b^2(\tau)\sin^2\left[2\pi f(t_0+(i-1)\Delta t_\mathrm{M}+\tau)\right]}{p}}_{Term~4} + \underbrace{\frac{\sum_{i=1}^{i = p} b^2(\tau)B^2_{g_1}{\mathcal{R}}_1^2(t_0+(i-1)\Delta t_\mathrm{M}+\tau)}{p}}_{Term~5}+ \\ \underbrace{\frac{\sum_{i=1}^{i = p} 2B_{g_2}\sin(2\pi f(t_0+(i-1)\Delta t_\mathrm{M})) \mathcal{R}_2(t_0+(i-1)\Delta t_\mathrm{M})}{p}}_{Term~6} - \underbrace{\frac{\sum_{i=1}^{i = p} 2a(\tau)\sin(2\pi f(t_0+(i-1)\Delta t_\mathrm{M}))}{p}}_{Term~7}+ \\ -\underbrace{\frac{\sum_{i=1}^{i = p} 2b(\tau)\sin(2\pi f(t_0+(i-1)\Delta t_\mathrm{M}))\sin(2\pi f(t_0+(i-1)\Delta t_\mathrm{M}+\tau))}{p}}_{Term~8} + \\ -\underbrace{\frac{\sum_{i=1}^{i = p} 2b(\tau)B_{g_1}\sin(2\pi f(t_0+(i-1)\Delta t_\mathrm{M}))\mathcal{R}_1(t_0+(i-1)\Delta t_\mathrm{M}+\tau)}{p}}_{Term~9}+ \\ -\underbrace{\frac{\sum_{i=1}^{i = p} 2a(\tau)B_{g_2}\mathcal{R}_2(t_0+(i-1)\Delta t_\mathrm{M})}{p}}_{Term~10}+ \\ -\underbrace{\frac{\sum_{i=1}^{i = p} 2b(\tau)B_{g_2}\mathcal{R}_2(t_0+(i-1)\Delta t_\mathrm{M})\sin(2\pi f(t_0+(i-1)\Delta t_\mathrm{M}+\tau))}{p}}_{Term~11}+ \\ -\underbrace{\frac{\sum_{i=1}^{i = p} 2b(\tau)B_{g_1}B_{g_2}\mathcal{R}_2(t_0+(i-1)\Delta t_\mathrm{M})\mathcal{R}_1(t_0+(i-1)\Delta t_\mathrm{M}+\tau)}{p}}_{Term~12}+ \\ \underbrace{\frac{\sum_{i=1}^{i = p} 2a(\tau)b(\tau)\sin(2\pi f(t_0+(i-1)\Delta t_\mathrm{M}+\tau))}{p}}_{Term~13}+ \underbrace{\frac{\sum_{i=1}^{i = p} 2a(\tau)b(\tau)B_{g_1}\mathcal{R}_1(t_0+(i-1)\Delta t_\mathrm{M}+\tau)}{p}}_{Term~14}+ \\ \underbrace{\frac{\sum_{i=1}^{i = p} 2b^2(\tau)B_{g_1}\sin(2\pi f(t_0+(i-1)\Delta t_\mathrm{M}+\tau))\mathcal{R}_1(t_0+(i-1)\Delta t_\mathrm{M}+\tau)}{p}}_{Term~15}.
\end{multline}
In the present study, $t_0=0.025$~s and $\Delta t_\mathrm{M}=5$~s for both test conditions. Using these and for $p \rightarrow \infty$, it can be shown that Terms~1--5 in Eq.~(\mbox{\ref{eq:RMSE3}}) equal $1/2$, $B_{g_2}^2/3$, $a^2(\tau)$, $b^2(\tau)/2$, and $b^2(\tau) B_{g_1}^2/3$, respectively. Also, Terms~6, 7, and 9--15 are zero for $p \rightarrow \infty$. Using trigonometric relations, it can be shown that Term~8 reduces to $-b(\tau)\cos(2 \pi f\tau)$. As a result, for an infinitely large number of training datasets, it can be obtained that
\begin{equation}
\label{eq:RMSE4}
RMSE^2(\tau,\infty) = \left(\frac{1}{2}+\frac{B^2_{g_1}}{3}\right)b^2(\tau)-b(\tau)\cos(2 \pi f \tau)+a^2(\tau)+\frac{B^2_{g_2}}{3}+\frac{1}{2}.
\end{equation}
Analyses presented in section~\ref{framework} showed that, for $B_{g_1}$ and $B_{g_2}$ smaller than or equal to unity, $a(\tau)$ is close to zero and nearly independent of $\tau$. This can be seen for the results presented in Fig.~\ref{fig:3Dfits}(a), with $B_{g_1} = B_{g_2} = 0.5$. Also, by definition, the slope of the best fit at a given value of $\tau$, i.e. $b(\tau)$, is estimated so that $RMSE$ is minimized. Thus, $b(\tau)$ can be obtained using $\partial RMSE/\partial b = 0$, which leads to
\begin{equation}
\label{eq:RMSE5}
b(\tau) = \frac{\cos(2\pi f \tau)}{1+\frac{2B^2_{g_1}}{3}}.
\end{equation}
Substituting $b(\tau)$ from Eq.~(\ref{eq:RMSE5}) into Eq.~(\ref{eq:RMSE4}), a closed form for $RMSE(\tau,\infty)$ is obtained and is given by
\begin{equation}
\label{eq:RMSE6}
RMSE(\tau,\infty) = \sqrt{\frac{1}{2}+\frac{B^2_{g_2}}{3}-\frac{\cos^2(2\pi f \tau)}{2+\frac{4B^2_{g_1}}{3}}}.
\end{equation}
The predictions of Eq.~(\ref{eq:RMSE6}) are presented in Fig.~\ref{fig:error} using the black dotted-dashed curves.

\bibliography{spray}

\end{document}